\documentclass{elsarticle}
\usepackage[utf8]{inputenc}
\usepackage[top=1in,bottom=1in,left=1in,right=1in]{geometry}
\usepackage{amsmath,amssymb}
\numberwithin{equation}{section}
\numberwithin{figure}{section}
\usepackage{xcolor}
\usepackage{subcaption,graphicx}
\usepackage{tikz}
\usepackage[ruled,vlined]{algorithm2e}

\graphicspath{{figures/}}

\newcommand{\bx}{\mathbf{x}}

\numberwithin{equation}{section}
\numberwithin{figure}{section}
\numberwithin{table}{section}
\newcommand{\vk}{{\bf k}}
\newcommand{\vq}{{\bf q}}

\begin{document}
\begin{frontmatter}

\title{Using Dynamic Model Decomposition to Predict the Dynamics of a Two-time Non-equilibrium Green's function}

\author[1]{Jia {Yin}\corref{cor1}}
\ead{jiayin@lbl.gov}
\author[2]{Yang-hao {Chan}}
\author[3]{Felipe da {Jornada}}
\author[4]{Diana {Qiu}}
\author[5,6]{Steven G. {Louie}}
\ead{sglouie@berkeley.edu}
\author[1]{Chao {Yang}\corref{cor1}}
\ead{cyang@lbl.gov}
\cortext[cor1]{Corresponding author:
	Tel.: +1-925-285-2235 (Jia Yin), +1-510-486-6424 (Chao Yang);  
	fax: +1-510-486-5812 (Chao Yang)}

\address[1]{Applied Mathematics \& Computational Research Division, Lawrence Berkeley National Laboratory, Berkeley, CA 94720, USA}
\address[2]{Institute of Atomic and Molecular Sciences, Academia Sinica, Taipei 10617, Taiwan}
\address[3]{Department of Materials Science and Engineering, Stanford University, Stanford, CA 94305, USA}
\address[4]{School of Engineering \& Applied Science, Yale University, New Haven, CT 06520, USA}
\address[5]{Department of Physics, University of California at Berkeley, Berkeley, CA 94720, USA}
\address[6]{Materials Sciences Division, Lawrence Berkeley National Laboratory, Berkeley, CA 94720, USA}

\begin{abstract}

Computing the numerical solution of the Kadanoff-Baym equations, a set of nonlinear integral differential equations satisfied by two-time Green's functions derived from many-body perturbation theory for a quantum many-body system away from equilibrium, is a challenging task. Recently, we have successfully applied dynamic mode decomposition (DMD) to construct a data driven reduced order model that can be used to extrapolate the time-diagonal of a two-time Green's function from numerical solution of the KBE within a small time window. 
In this paper, we extend the previous work and use DMD to predict off-diagonal elements of the two-time Green's function. We partition the two-time Green's function into a number of one-time functions along the diagonal and subdiagonls of the two-time window as well as in horizontal and vertical directions. We use DMD to construct separate reduced order models to predict the dynamics of these one-time functions in a two-step procedure. We extrapolate along diagonal and several subdiagonals within a subdiagonal band of a two-time window in the first step. In the second step, we use DMD to extrapolate the Green's function outside of the sub-diagonal band. We demonstrate the efficiency and accuracy of this approach by applying it to a two-band Hubbard model problem.
\end{abstract}


\end{frontmatter}

\section{Introduction} \label{sec:intro}

%
Simulating a quantum many-body system away from equilibrium is a challenging
task. Although time-dependent physical observables can be computed from the solution of a time-dependent Schr\"{o}dinger equation with a time-dependent Hamiltonian, such a brute-force approach is limited to small systems defined in a small dimensional Hilbert space. A more practical approach is to focus on the Green's function, $G(t,t')$, which is a two-point correlator of the creation and annhilation field operators defined on the Keldysh contour~\cite{BK,kadanoff2018quantum,keldysh1965zhetf,FW,lipavsky1986generalized,Hermanns2012}. Unlike the equilibrium Green's function, which depends on $t-t'$, hence is a one time function, the non-equilibrium Green's function depends on both $t$ and $t'$. The equation of motion satisfied by the two-time non-equilibrium Green's function is a set of nonlinear integro-differential equations of the form
\begin{equation}\label{eq:kbe}
    \left[i\frac{d}{dt}-H(t)\right]G(t, t') = \delta(t, t') + \int_C\Sigma(t, \overline{t})G(\overline{t}, t')d\overline{t},
\end{equation}
where $H(t)$ is a single-particle Hamiltonian, $\Sigma(t,t')$ is a self-energy that accounts for the many-body interactions that can depend on the Green's function $G$.
Equation \eqref{eq:kbe} and its accompanying adjoint equation which describes the propagation of $G(t,t')$ along the $t'$ direction are often referred to as the Kadanoff-Baym equations (KBE)~\cite{kadanoff2018quantum}.

Evolving the Green's function numerically on a two-time grid is highly non-trivial. A commonly used method to solve the KBE system~\eqref{eq:kbe} is a nonlinear time evolution scheme based on the second-order implicit Runge-Kutta method and a fixed point iteration~\cite{Stan2009}.  Other explicit/implicit Runge-Kutta schemes can also be applied to solve the ODE system~\cite{hoppensteadt2007numerical,kennedy2016diagonally}. In these numerical methods, the presence of the integral kernel in the KBEs makes both the memory requirement and computational cost high if the long-time behavior of a physical observable is to be examined~\cite{burton2005volterra}. 

In~\cite{DMDdiag}, we proposed using a model reduction technique called the dynamic mode decomposition (DMD) method, to predict the time diagonal of the Green's function, i.e, $\rho(t)=G(t,t)$, for large $t$, from
a linear reduced order model constructed from the numerical solution of the KBE within a small two-time window. Even though the two-time dynamics satisfied by $G(t,t')$ can be nonlinear,  we observed that the one-time dynamics satisfied by $\rho(t)$, which cannot be easily written down analytically, can be well approximated by a linear model.

In this paper, we extend the technique developed in~\cite{DMDdiag} to use DMD to predict the entire two-time Green's function $G(t,t')$ from the numerical solution of the KBE from a small time window. Our basic
strategy is to divide the two-time $G(t,t')$ into a number of one-time functions and construct a DMD-based reduced order model for each one of them.  We examine a few different ways to perform such divisions and approximations. In one approach, we treat $G(t,t')$ for a fixed $t-t'$ as a one-time function that satisfies a one-time dynamical system defined by the KBE. The prediction of such time subdiagonal of $G(t,t')$ is a 
natural extension of the prediction of the time diagonal of $G(t,t')$ we developed in the previous work.
After the prediction of several time-subdiagonals of $G(t,t')$ have been made, we can then use either the computed or extrapolated $G(t,t')$ for a fixed $t'$ to construct a reduce order model to predict the values of $G(t,t')$ for large $t'$s, i.e. away from the time diagonal. This two-step procedure is compared with an alternative approach in which we first fix $t'$ and extrapolate along the $t$ direction, and then fix $t$ and extrapolate along the $t'$ direction. The DMD method employed in both of these two approaches provides a momentum-temporal decomposition of several one time functions. Numerical examples are presented to demonstrate the effectiveness of the DMD extrapolation for a simple Hubbard model driven by an external field with different intensity levels. Yet another alternative we consider in this paper is a decomposition that fixes the $k$-point (i.e., a momentum grid point) and treats one of the time variable as a spatial variable within a selected time window. The reduced order model constructed in this scheme allows us to extrapolate the values of $G(t,t')$ within a two-time sub-window for a specific $k$-point. We show, by numerical examples, that this approach can sometimes be more effective than a momentum-temporal decomposition.

The rest of the paper is organized as follows. In Section~\ref{sec:dmd}, we review the principal ideas and procedures of applying DMD. The implementation of DMD for the two-time Green's function is discussed in Section~\ref{sec:method}. In Section~\ref{sec:example}, we demonstrate the effectiveness of the proposed DMD schemes by  numerical examples.

\section{Dynamic mode decomposition} \label{sec:dmd}

In this section, we provide an overview of the dynamic mode decomposition (DMD) method to be used in the next
section to predict values of the two-time Green's function $G(t,t')$ for large $t$ and $t'$ from the numerical solution of the KBE within a small two-time window. We will also discuss a variant of DMD called high order DMD (HODMD) that can yield more accurate prediction from spatially undersampled data.

DMD is a data-driven dimension reduction technique used to construct a low dimensional linear
dynamical model that can be used to predict observables of a nonlinear dynamical system with a large number of degrees of freedom~\cite{kutz2016dynamic,DMD0,schmid2011applications,TuRowley}.  The linear model 
can be characterized by a number of spatial and temporal modes that can be obtained from the eigenvalues and eigenvectors of a linear operator.

To describe the basic idea of DMD, consider a nonlinear dynamical system defined by the ordinary differential equation
\begin{equation}\label{eq:model}
    \frac{d\mathbf{x}(t)}{dt} = \mathbf{f}(\mathbf{x}(t), t), \quad t\geq 0,
\end{equation}
where $\mathbf{x}(t)\in\mathbb{C}^n$ is a time-dependent state variable, and $\mathbf{f}: \mathbb{C}^n\otimes \mathbb{R}^+ \rightarrow \mathbb{C}^n$ is a nonlinear function of $\mathbf{x}$ and time $t$. 

If we were to approximate \eqref{eq:model} by a linear model
\begin{equation}
    \frac{d\mathbf{x}(t)}{dt} = \mathbf{A} \mathbf{x}(t),
\end{equation}
what is the best choice of the linear operator $\mathbf{A}$? This question is important for problems in which distinct features such as certain oscillation frequencies and amplitude decay rate are of interest even though the overall dynamics cannot be easily described by a linear model. For problems that has 
an explicit analytical expression of $\mathbf{f}(\mathbf{x}(t), t)$, it may be possible to linearize $\mathbf{f}(\mathbf{x}(t), t)$ and derive $\mathbf{A}$ explicitly. This linearization process essentially amounts to a linear response analysis.  However, when the analytical form of  $\mathbf{f}(\mathbf{x}(t), t)$ is unknown, performing such an analysis is difficult, if not impossible.

The linearization produced by DMD is based the Koopman operator theory~\cite{DMDtoKoop,Koopman1,Koopman2}, which is developed to characterize the evolution from a scalar obervable function of $\mathbf{x}(t)$, denoted by $g(\mathbf{x}(t))$, to $g(\mathbf{x}(t+\Delta t))$, i.e.
\[
g(\mathbf{x}(t+\Delta t)) = \mathcal{K}_{\Delta t} g(\mathbf{x}(t)).
\]

In the limit of $\Delta t \rightarrow 0$, the Koopman operator defines a linear dynamical system
\[
\frac{dg(\mathbf{x}(t))}{dt} = \mathcal{K} g(\mathbf{x}(t)).
\]

Because the Koopman operator $\mathcal{K}$ is a linear operator that maps from a function space to 
another function space, it has infinite number of eigenvalues $\lambda_j$ and eigenfunctions 
$\varphi_j(\mathbf{x})$, $j = 1, 2, ..., \infty$. 

If the observable functions of interest can be well approximated by an invariant subspace of 
$\mathcal{K}$ defined by a finite subset of eigenvalues and eigenvectors, then it is possible to 
construct a finite dimensional operator (matrix) approximation to $\mathcal{K}$.

To be specific, if $g_1(\mathbf{x})$, $g_2(\mathbf{x})$,...,$g_n(\mathbf{x})$ 
are $n$ observable functions that can be expressed as
\[
\begin{bmatrix}
g_1(\mathbf{x}) \\
g_2(\mathbf{x}) \\
\vdots \\
g_n(\mathbf{x}) 
\end{bmatrix}
 = 
\begin{bmatrix}
v_1 & v_2 & \cdots & v_k
\end{bmatrix}
\begin{bmatrix}
\varphi_1(\mathbf{x}) \\
\varphi_2(\mathbf{x}) \\ 
\vdots \\
\varphi_k(\mathbf{x}),
\end{bmatrix}
\]
for some vectors $v_1, v_2, ..., v_k \in \mathbb{C}^n$, which contain the expansion coefficients, 
then $\mathcal{K}$ can be approximated by a $k \times k$ matrix $\mathbf{A}$.


To construct such an approximation for observable functions that are chosen to be the components of
$\mathbf{x}(t)$ defined in \eqref{eq:model}, we take snapshots of $\mathbf{x}(t)$ at $t_j = (j-1)\Delta t$, i.e., $\mathbf{x}_j = \mathbf{x}(t_j)$, for $j=1,...,m$, and use them to build two matrices 
$\mathbf{X}_1$ and $\mathbf{X}_2$ of the form
\begin{equation}
\mathbf{X}_1 =\left( \mathbf{x}_1 \: \mathbf{x}_2 \: \cdots \: \mathbf{x}_{m-1} \right) \ \ \mbox{and} \ \
\mathbf{X}_2 =\left( \mathbf{x}_2 \: \mathbf{x}_3 \: \cdots \: \mathbf{x}_{m} \right).
\label{eq:mats}
\end{equation}
The finite dimensional approximation to the Koopman operator can then be obtained by solving the following linear least squares problem
\begin{equation}
 \min_{\mathbf{A}} \| \mathbf{A} \mathbf{X}_1 - \mathbf{X}_2 \|_F^2.
\label{eq:lsq}
\end{equation}
The solution to \eqref{eq:lsq}  is
\begin{equation}
    \mathbf{A} = \mathbf{X}_2\mathbf{X}_1^\dagger,
\end{equation}
where $\mathbf{X}_1^\dagger$ is the Moore-Penrose pseudoinverse of $\mathbf{X}_1$ that can be computed from the singular value decomposition (SVD)~\cite{SVD} of $\mathbf{X}_1$.
If the nonzero singular values of $\mathbf{X}_1$, $\sigma_j$, $j = 1,2,...,m$, decrease rapidly with respect to $j$, which indicates that the numerical rank, denoted by $r$, of $\mathbf{X}_1$ is much smaller than $m$ and $n$, then $\mathbf{A}$ can be further approximated by a truncated SVD, 
\begin{equation}
\mathbf{A} \approx \mathbf{X}_2\widetilde{\mathbf{V}}\widetilde{\mathbf{\Sigma}}^{-1}\widetilde{\mathbf{U}}^*,
\label{eq:projA}
\end{equation}
where the $r\times r$ diagonal matrix $\widetilde{\mathbf{\Sigma}}$ contains the leading $r$ dominant 
singular values of $\mathbf{X}_1$, and $\widetilde{\mathbf{U}}$ and $\widetilde{\mathbf{V}}$ 
contain the corresponding right and left singular vectors of $\mathbf{X}_1$.


We can now fully characterize the approximated reduced order linear dynamical system model by 
diagonalizing the projected Koopman operator
$\widetilde{\mathbf{A}} = \widetilde{\mathbf{U}}^\ast \mathbf{A} \widetilde{\mathbf{U}} 
= \widetilde{\mathbf{U}} \mathbf{X}_2 \widetilde{\mathbf{V}} \widetilde{\mathbf{\Sigma}}^{-1}$.
Let 
\begin{equation}
\widetilde{\mathbf{A}}\mathbf{W} = \mathbf{W}\mathbf{\Lambda},
\label{eq:dmdev}
\end{equation}
be the eigendecomposition of $\widetilde{\mathbf{A}}$, 
where $\mathbf{\Lambda} = {\rm{diag}}(\lambda_1, ..., \lambda_r)$ is composed of the eigenvalues of $\widetilde{\mathbf{A}}$, and the columns of $\mathbf{W}$ are the corresponding eigenvectors.
The matrix
\begin{equation}
\mathbf{\Phi} = \mathbf{X}_2\widetilde{\mathbf{V}}\widetilde{\mathbf{\Sigma}}^{-1}\mathbf{W}
\label{eq:dmdmodes}
\end{equation}
contains the so called the DMD modes. If $\phi_{\ell}$ is the $\ell$th column of $\mathbf{\Phi}$,
the DMD approximation to $\mathbf{x}$ can be by represented by
\begin{equation}\label{eq:evol_DMD}
\mathbf{x}(t)\approx \sum_{\ell=1}^r\mathbf{\phi}_\ell\exp(i\omega_\ell^{\text{DMD}} t)b_\ell = \mathbf{\Phi}\exp(\Omega t)\mathbf{b}.
\end{equation}
where $\omega_\ell^{\text{DMD}} = -i\frac{\ln{\lambda_\ell}}{\Delta t}$, $\ell = 1, ..., r$, $\mathbf{\Omega} = \frac{\ln{\mathbf{\Lambda}}}{\Delta t} = {\rm{diag}}(i\omega_1^{\rm{DMD}}, ..., i\omega_r^{\rm{DMD}})$, and the amplitude vector $\mathbf{b}:=[b_1, ..., b_r]^T$ is taken either as the projection of the initial value $\mathbf{x}_1$ onto the DMD modes, i.e.,
\begin{equation}\label{eq:b1}
\mathbf{b} = \mathbf{\Phi}^\dagger \mathbf{x}_1,
\end{equation}
or as the least squares fit of \eqref{eq:evol_DMD} on the sampled trajectories, i.e.,
\begin{equation}\label{eq:b2}
\mathbf{b} = \arg\min_{\tilde{\mathbf{b}}\in\mathbb{C}^n}\sum_{j=1}^m\|\mathbf{\Phi}\exp(\Omega t_j)\tilde{\mathbf{b}}-\bx_j\|^2,
\end{equation}
where  $\|\cdot\|$ denotes the standard Euclidean norm of a vector. For more details on the numerical procedure, we refer to~\cite{kutz2016dynamic,DMD0,TuRowley} and paper~\cite{DMDdiag}.

The major computational cost of DMD computation is in the SVD of $\mathbf{X}_1$, which is $O(\min(m^2n, mn^2))$. The memory cost is $O(mn)$.

As pointed out in~\cite{TuRowley,wkr15}, the success of the DMD approximation to the Koopman operator depends crucially on the choice of observables. When the observables are chosen to be discretized components of $\mathbf{x}$, a limited resolution in the discretization may lead to a poor DMD approximation to the Koopman operator as shown in~\cite{DMDdiag}. In particular, the number of DMD modes $r$,  which can be extracted from the data, may be too small to represent the true dynamics of $\bx(t)$. To resolve this problem, we can use the higher order DMD (HODMD) method, which can be derived from the time-delay embedding theory~\cite{broomhead1986extracting,packard1980geometry,Pan2020,Taken}. For more details about this theory and its relation to HODMD, we refer readers to~\cite{HODMD} and~\cite{DMDdiag}.

In HODMD($d$), each column of $\mathbf{X}_1$ and $\mathbf{X}_2$ consists of $d$ consecutive snapshots. The same snapshot may be used in several adjacent columns. This construction increases the leading dimension of the data matrices by a factor of $d$.  Consequently, the cost of computing the HODMD modes is also higher.  Furthermore, when $\Delta t$ is small, the columns can become more linearly dependent. To reduce the computational cost and column linear dependency of the data matrix, we can increase the temporal distance between the augmented snapshots to make sure there is no overlapping between two columns of $\mathbf{X}_1$ and $\mathbf{X}_2$. Specifically, we can define the data matrices $\widetilde{\mathbf{X}}_1$ and $\widetilde{\mathbf{X}}_2$ as
\begin{equation}\label{eq:DMDc-d1}
	\widetilde{\mathbf{X}}_1 = \begin{bmatrix}
		\mathbf{x}_1 & \mathbf{x}_{d+1} & ... & \mathbf{x}_{(p-2)d+1}\\
		\mathbf{x}_2 & \mathbf{x}_{d+2} & ... & \mathbf{x}_{(p-2)d+2}\\
		\vdots & \vdots & \vdots & \vdots\\
		\mathbf{x}_{d} & \mathbf{x}_{2d} & ... & \mathbf{x}_{(p-1)d}
	\end{bmatrix}, \quad
	\widetilde{\mathbf{X}}_2 = \begin{bmatrix}
		\mathbf{x}_{d+1} & \mathbf{x}_{2d+1} & ... & \mathbf{x}_{(p-1)d+1}\\
		\mathbf{x}_{d+2} & \mathbf{x}_{2d+2} & ... & \mathbf{x}_{(p-1)d+2}\\
		\vdots & \vdots & \vdots & \vdots\\
		\mathbf{x}_{2d} & \mathbf{x}_{3d} & ... & \mathbf{x}_{pd}
	\end{bmatrix},
\end{equation}
where $p=\text{floor}(m/d)$. 

Once $\widetilde{\mathbf{X}}_1$ and $\widetilde{\mathbf{X}}_2$ are prepared according to \eqref{eq:DMDc-d1},
we follow the same procedure used in DMD to compute the HODMD modes. The only difference is that the time step between two adjacent columns becomes $d\times\Delta t$ instead of $\Delta t$. As each column of $\widetilde{\mathbf{X}}_1$ consists of $d$ consecutive snapshots, each spatial HODMD mode is a vector of length $nd$. Therefore, in the reconstruction and extrapolation of $\mathbf{x}(t)$ by \eqref{eq:evol_DMD}, we only take the first $n$ elements of each spatial HODMD mode as $\phi_{\ell}$, $\ell = 1, ..., r$.

\section{DMD for the two-time Green's function} \label{sec:method}


\subsection{DMD of $G(t,t')$ for a fixed $t-t'$ or $t'$} 
\label{sec:fixt}
In previous work~\cite{DMDdiag}, we applied the DMD technique to analyze and 
extrapolate the time diagonal of the two-time Green's function $G(t,t')$
from the numerical solution of the KBE \eqref{eq:kbe} within a small
time window. 

To be specific, we solved the KBE within the time window
$[0,t_m]\times [0,t_m]$ for a sufficiently small 
time step $\Delta t$ and a small integer $m$. The time diagonal
of the Green's function $\rho(k_s;t_j) = G(k_s;(j-1)\Delta t,(j-1)\Delta t)$, with $s=1, \,..., \,n_k$, $j = 1,\, ...,\, m$ were used to
construct a snapshot matrix $\mathbf{X}$ as
\begin{equation}
\mathbf{X} = \left[
\rho(\mathbf{k}, t_1), \; \rho(\mathbf{k}, t_2), \; ..., \; \rho(\mathbf{k}, t_m)
\right],
\end{equation}
where $\mathbf{k}:= (k_1,\, ...,\, k_{n_k})^T$ denotes the uniformly sampled $k$-points in the Brillouin zone of the
momentum space, and each snapshot $\rho(\mathbf{k}, t_j)$ is defined as
\begin{equation}
    \rho(\mathbf{k}, t_j) = \left(\rho(k_1, t_j),\, \rho(k_2, t_j),\, ...,\, \rho(k_{n_k}, t_j)\right)^T, \quad j = 1, ..., m.
\end{equation}

We have shown that, for a two-band Hubbard model driven by an external
field, DMD can successfully predict the long time dynamics of $\rho$
from $\mathbf{X}$ when the intensity of the driving field is relatively
small.  For high intensity driving field, HODMD can be used to accurately
predict the long time dynamics of $\rho$ from the solution of KBE within a
small time window.

The success of DMD and HODMD is partly due to the fact that the time diagonal
of $G(t,t')$ is well behaved, i.e., the real and imaginary parts of this function
are smooth, and they exhibit clear oscillation and decay properties.

\begin{figure}[t!]
    \centering
    \includegraphics[width=0.9\textwidth]{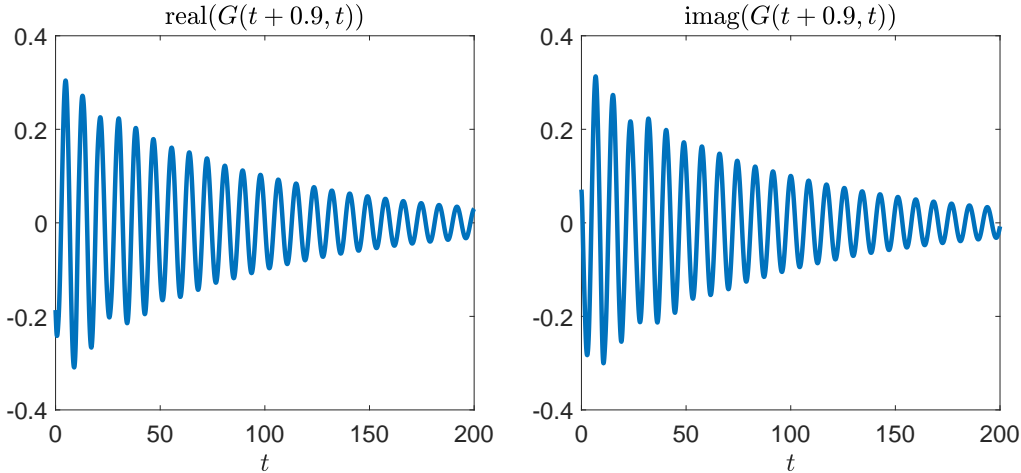}
    \caption{The subdiagonal $G(t+0.9, t)$ for the second Born model with the intensity of the external field $I=0.5$ at $k=0$.}
    \label{fig:Gsubdiag}
\end{figure}

It has been observed that the smoothness property of $G(t,t')$ also holds
for $t'-t=\tau$, where $\tau > 0$ is fixed. As an example, Figure~\ref{fig:Gsubdiag} shows $G(t+0.9,t)$, i.e., $\tau = 0.9$, for the same two-band Hubbard model examined in~\cite{DMDdiag}.
Therefore, we can, in principle, use the same DMD and HODMD techniques we developed for 
predicting the time diagonal of $G$ to predict the time subdiagonals of 
$G$. 

\begin{figure}[t!]
\centering
\begin{tikzpicture}
    \draw[->] (0,0)--(5,0);
    \draw[->] (0,0)--(0,5);
	\draw (0,0) -- (0,0.1);
	\draw (1.2,0) -- (1.2,0.1);
	\draw (2.2,0) -- (2.2,0.1);
	\draw (4.7,0) -- (4.7,0.1);
	\draw (0,0) -- (0.1,0);
	\draw (0,1) -- (0.1,1);
	\draw (0,2.2) -- (0.1,2.2);
	\draw (0, 4.7) -- (0.1,4.7);
	
    \node[below] at (-0.2,0){0};
    
    \node[below] at (0.55, -0.12){$\dots$};
    \node[below] at (1.2, 0){$t_{\ell}$};
    \node[below] at (2.2, 0){$t_{m+\ell-1}$};
    \node[below] at (3.1, -0.12){$\dots$};
    \node[below] at (4.7, 0){$t_{N}$};
    \node[right] at (5.1, 0){$t$};
    	
    \node[left] at (-0.12, 0.5){$\vdots$};		
    \node[left] at (0, 1){$t_{m}$};
    \node[left] at (0, 2.2){$t_{m+\ell-1}$};
    \node[left] at (-0.12, 3.5){$\vdots$};
    \node[left] at (0, 4.7){$t_{N}$};
    \node[left] at (0, 5.1){$t'$};
    			
    \coordinate (O) at (0, 0);
    \coordinate (P1) at (1.2, 0);
    \coordinate (P2) at (2.2, 1);
    \coordinate (P4) at (4.7, 4.7);
    \coordinate (P5) at (2.2, 0);
    \coordinate (P6) at (0, 2.2);
    \coordinate (P7) at (2.2, 2.2);
    \coordinate (P8) at (0, 4.7);
    \coordinate (P9) at (0, 1);
    \coordinate (P10) at (4.7, 0);
    \coordinate (P11) at (4.7, 3.5);
    \coordinate (P12) at (2.6, 2);
    \coordinate (P13) at (4.2, 3.6);

    \draw[thick, color=red] (O) -- (P5) -- (P7) -- (P6) -- cycle;
    \draw[thick, color=blue, fill=gray!30] (O) -- (P1) -- (P2) -- (P7) -- cycle;
    \draw[thick, dashed, color=blue] (P7) -- (P4) -- (P11) -- (P2) -- cycle;
    \draw[thick, dashed] (P4) -- (P8);
    \draw[thick, dashed] (P2) -- (P9);
    \draw[thick, dashed] (P4) -- (P10);
    
    \draw[->, dashed, thick, blue] (P12) -- (P13);
    
    \fill[red] (O) circle (1.5pt);
    \fill[red] (P5) circle (1.5pt);
    \fill[red] (P6) circle (1.5pt);
    \fill[red] (P7) circle (1.5pt);
    \fill[blue] (P1) circle (1.5pt);
    \fill[blue] (P2) circle (1.5pt);
\end{tikzpicture}

\caption{To use DMD or HODMD to predict $G(t,t')$ along the time diagonal and $\ell-1$ time subdiagonals (contained in the parallegram outlined by the blue dashed lines), we need to solve the KBE numerically within the time window $[0,t_{m+\ell-1}] \times [0,t_{m+\ell-1}]$ (drawn in red). The snapshot matrices are constructed by extracting subdiagonals of $G(t,t')$ within the shaded parallelogram contained in the solid blue lines.}
\label{fig:subwindow}
\end{figure}
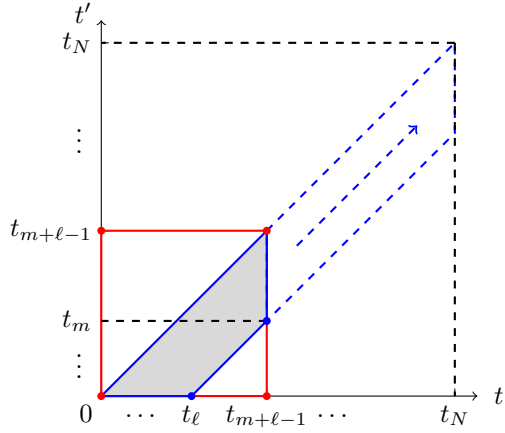

However, to generate a snapshot matrix $\mathbf{X}$ for the $j$-th subdiagonal of $G$ with $m$ snapshots, we need to increase the size of the sampled time window in which the KBE is solved numerically from $[0,t_m] \times [0,t_m]$ 
to $[0,t_{m+j}] \times [0,t_{m+j}]$. The $n_k\times m$ snapshot matrix $\mathbf{X}$ for the $j$-th subdiagonal $G(\mathbf{k}; t+j\Delta t, t)$ is therefore given by
\begin{equation}\label{eq:md1_dm1}
    \mathbf{X} = \left[
    G(\mathbf{k}, t_{1+j}, t_1), \; G(\mathbf{k}, t_{2+j}, t_2), \; ..., \; G(\mathbf{k}, t_{m+j}, t_m)
    \right].
\end{equation}
In general, if the diagonal and $\ell-1$ subdiagonals of $G(t,t')$ are to be analyzed and extrapolated,
we need to solve the KBE within the time window of $[0,t_{m+\ell-1}] \times [0,t_{m+\ell-1}]$.
Figure~\ref{fig:subwindow} gives a schematic depiction of the time window
in which the KBE is solved. The shaded region contains the snapshots to be used for DMD or HODMD analysis. The extrapolated subdiagonal elements of $G(t,t')$ are contained in the parallelgram denoted by the blue dashed lines.  
Note that, within the time window $[0,t_{m+\ell-1}] \times [0,t_{m+\ell-1}]$, more snapshots can be used to perform DMD or HODMD for the subdiagonals closer to the diagonal.

\begin{figure}[t!]
    \centering
    \includegraphics[width=0.9\textwidth]{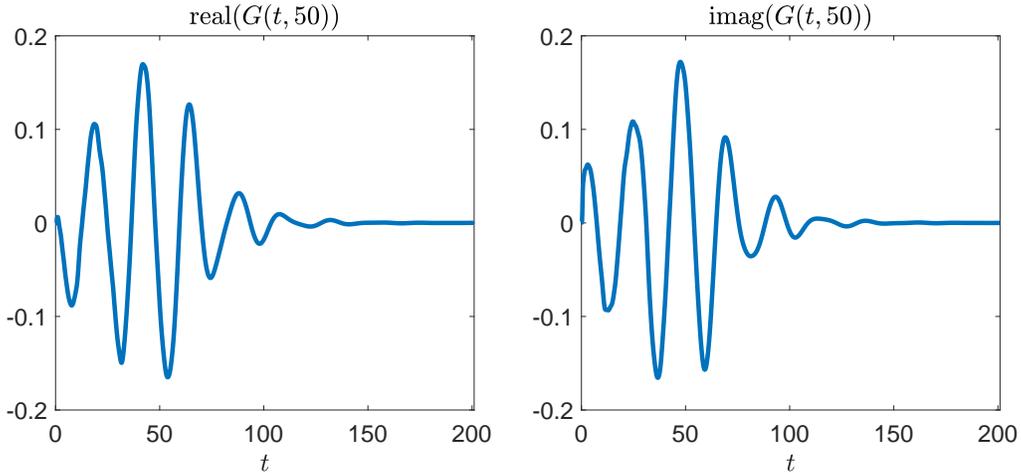}
    \caption{The real and imaginary parts of $G(t, 50)$ at $k = 0$ for the two-band Hubbard model with the intensity of the external field set to $I=0.5$.}
    \label{fig:Gtprime}
\end{figure}

To obtain values of $G(t,t')$ for $t$ and $t'$ that are outside of the 
subdiagonal band, we rely on another observation that shows $G(t,t')$ is typically smooth with respect to $t$ for a fixed $t'$ and vice versa. For example, In Figure~\ref{fig:Gtprime}, we plot the real and imaginary parts of $G(t, t')$ at $t'=50$. Both curves are smooth with clear oscillation frequencies and amplitude envelops.

Therefore, we can utilize values in the subdiagonal band to extrapolate those that are outside the band. To be specific, for a fixed $t'=t_j$, the $n_k\times \ell$ data matrix $\mathbf{X}$ to be used in DMD/HODMD is constructed as
\begin{equation}\label{eq:md1_dm2}
    \mathbf{X} = \left[
    G(\mathbf{k}, t_{j}, t_j), \; G(\mathbf{k}, t_{j+1}, t_j), \; ..., \; G(\mathbf{k}, t_{j+\ell-1}, t_j)
    \right].
\end{equation}
If $j\leq m$, then from Figure~\ref{fig:subwindow}, these data can be taken from the numerical solution to KBE. Otherwise, they are approximated by subdiagonal extrapolations produced in the previous steps.

What we have described is a two-step DMD method for  predicting $G(t,t')$ for large $t$ and $t'$ from the numerical solution of the KBE within a small two-time window.  In the first step, the values in the subdiagonal band $G(t+j\Delta t, t)$ with $j=0, 1, ..., \ell-1$ are extrapolated by applying DMD to the data matrices \eqref{eq:md1_dm1}. In the second step, we use the solutions in the subdiagonal band to predict the values of $G(t,t')$ outside the subdiagonal band. As in each step, one time direction is always fixed ($t-t'$ for the first step, and $t'$ for the second step), we call this method fixed timeline (FT) DMD extrapolation.

\subsection{DMD of $G(t,t')$ for a fixed $k$ point}
\label{sec:fixk}

DMD is intrinsically a two-dimensional decomposition technique. In a traditional application of DMD, one of the dimension is the spatial (or momentum) dimension and the other is the temporal dimension. That is why DMD is known as the spatial-temporal decomposition of nonlinear dynamics.

However, the two-time Green's function used to model the dynamics of a many-body system away from equilibrium is a three dimensional tensor with momentum being one of the dimensions and $t$ and $t'$ being the other two. As a result, in order to use DMD, we must fix one of the dimensions and apply DMD to decompose $G$ in the other two dimensions.

In the previous section,  we took the more traditional approach by fixing either $t-t'$ or $t'$ and performing a DMD in the two dimensions defined by momentum ($k$) and time ($t$) for a fixed $t-t'$ or $t'$.

However, the DMD algorithm itself is agnostic to the physical interpretation of the dimensions and data.  All it requires are a few slices of data that are related and vary smoothly from one slice to another. These data slices can be combined and viewed as a data matrix on which a truncated singular value decomposition can be performed.  Once the DMD modes and frequencies are computed, they can be assembled to  construct a reduced order model for predicting additional data slices. 

In this section, we take an alternative approach in using DMD to analyze and predict the two-time Green's function.
Instead of fixing $t'$ or $t-t'$, we fix the $k$-point, and apply DMD directly to $G(t,t')$.

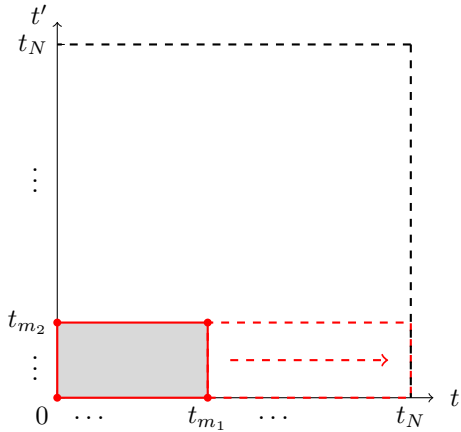
\begin{figure}[ht!]
\centering
\begin{tikzpicture}
    \draw[->] (0,0)--(5,0);
    \draw[->] (0,0)--(0,5);
	\draw (0,0) -- (0,0.1);
	\draw (2,0) -- (2,0.1);
	\draw (4.7,0) -- (4.7,0.1);
	\draw (0,0) -- (0.1,0);
	\draw (0,1) -- (0.1,1);
	\draw (0, 4.7) -- (0.1,4.7);
	
    \node[below] at (-0.2,0){0};
    
    \node[below] at (0.45, -0.12){$\dots$};
    \node[below] at (2, 0){$t_{m_1}$};
    \node[below] at (2.9, -0.12){$\dots$};
    \node[below] at (4.7, 0){$t_{N}$};
    \node[right] at (5.1, 0){$t$};
    	
    \node[left] at (-0.12, 0.5){$\vdots$};		
    \node[left] at (0, 1){$t_{m_2}$};
    \node[left] at (-0.12, 3){$\vdots$};
    \node[left] at (0, 4.7){$t_{N}$};
    \node[left] at (0, 5.1){$t'$};
    			
    \coordinate (O) at (0, 0);
    \coordinate (P1) at (4.7, 1);
    \coordinate (P2) at (2.3, 0.5);
    \coordinate (P3) at (4.4, 0.5);
    \coordinate (P4) at (4.7, 4.7);
    \coordinate (P5) at (2, 0);
    \coordinate (P6) at (0, 1);
    \coordinate (P7) at (2, 1);
    \coordinate (P8) at (0, 4.7);
    \coordinate (P9) at (0, 1);
    \coordinate (P10) at (4.7, 0);

    \draw[thick, color=red, fill = gray!30] (O) -- (P5) -- (P7) -- (P6) -- cycle;
    \draw[thick, dashed, color=red] (P7) -- (P1) -- (P10) -- (P5) -- cycle;
    \draw[thick, dashed] (P4) -- (P8);
    \draw[thick, dashed] (P4) -- (P10);
    
    \draw[->, dashed, thick, red] (P2) -- (P3);
    
    \fill[red] (O) circle (1.5pt);
    \fill[red] (P5) circle (1.5pt);
    \fill[red] (P6) circle (1.5pt);
    \fill[red] (P7) circle (1.5pt);
\end{tikzpicture}

\caption{Illustration of the extrapolation of $G(t, t')$ along the $t$ direction by applying DMD or HODMD to the sampled window $[0, t_{m_1}]\times [0, t_{m_2}]$ given by the shaded square.}
\label{fig:onedir}
\end{figure}

The simplest scheme is to take the numerical solution of the KBE within the time window $[0,t_{m_1}]\times [0,t_{m_2}]$ for a fixed $k$-point as the snapshot matrix, and perform DMD or HODMD to extrapolate and predict the values of $G(t,t')$ for $0\leq t' \leq t_{m_2}$ and $t>t_{m_1}$ as the first step. This procedure is illustrated in Figure~\ref{fig:onedir}. Then, in the second step, the computed or extrapolated values of $G(t,t')$ within $[0,t_N]\times [0,t_{m_2}]$ are used to predict values of $G(t,t')$ for $t' > t_{m_2}$.

Note that the KBE is typically solved 
for $t' \leq t$. The values of $G(t,t')$ in the upper triangular part of the time window $0 \leq t \leq t_{m_1}$ and $t'>t$ can be obtained from symmetry properties of $G$. However, because the amplitude envelop of $G(t,t')$ is typically different for $t < t'$ and $t\geq t'$ (for a fixed $t'$) (see Figure~\ref{fig:Gtprime}), using the values of $G(t,t')$ for $t \leq t'$ to predict $G(t,t')$ for $t > t' $ may not work well as we will show in the next section.

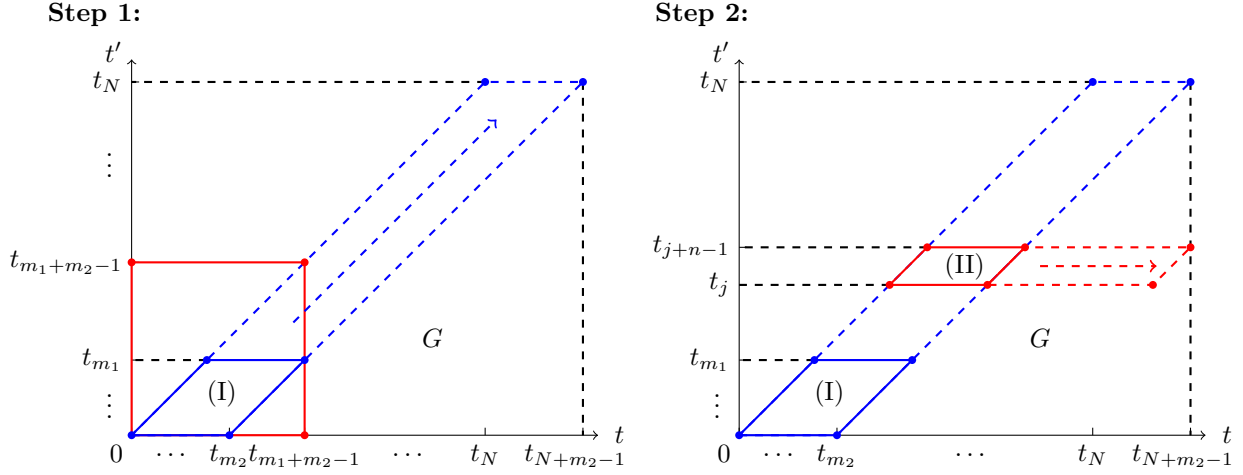
\begin{figure}[t!]
\centering
\begin{tikzpicture}
    \draw[->] (0,0)--(6.2,0);
    \draw[->] (0,0)--(0,5);
	\draw (0,0) -- (0,0.1);
	\draw (1.3,0) -- (1.3,0.1);
	\draw (2.3,0) -- (2.3,0.1);
	\draw (4.7,0) -- (4.7,0.1);
	\draw (6,0) -- (6,0.1);
	\draw (0,0) -- (0.1,0);
	\draw (0,1) -- (0.1,1);
	\draw (0,2.3) -- (0.1,2.3);
	\draw (0,4.7) -- (0.1,4.7);
	
    \node[below] at (-0.2,0){0};
    
    \node[below] at (0.55, -0.12){$\dots$};
    \node[below] at (1.3, 0){$t_{m_2}$};
    \node[below] at (2.3, 0){$t_{m_1+m_2-1}$};
    \node[below] at (3.7, -0.12){$\dots$};
    \node[below] at (4.7, 0){$t_{N}$};
    \node[below] at (5.9, 0){$t_{N+{m_2}-1}$};
    \node[right] at (6.3, 0){$t$};
    	
    \node[left] at (-0.12, 0.5){$\vdots$};		
    \node[left] at (0, 1){$t_{m_1}$};
    \node[left] at (0, 2.3){$t_{m_1+m_2-1}$};
    \node[left] at (-0.12, 3.7){$\vdots$};
    \node[left] at (0, 4.7){$t_{N}$};
    \node[left] at (0, 5.1){$t'$};
    \node[left] at (0.25, 5.6){\textbf{Step 1:}};
    			
    \coordinate (O) at (0, 0);
    \coordinate (P1) at (1.3, 0);
    \coordinate (P2) at (2.3, 1);
    \coordinate (P3) at (1, 1);
    \coordinate (P4) at (6, 4.7);
    \coordinate (P5) at (4.7, 4.7);
    \coordinate (P6) at (2.15, 1.5);
    \coordinate (P7) at (4.85, 4.2);
    \coordinate (P8) at (0, 4.7);
    \coordinate (P9) at (0, 1);
    \coordinate (P10) at (6, 0);
    \coordinate (P11) at (2.3, 0);
    \coordinate (P12) at (0, 2.3);
    \coordinate (P13) at (2.3, 2.3);

    \draw[thick, color=red] (O) -- (P11) -- (P13) -- (P12) -- cycle;
    \draw[thick, dashed, blue] (O) -- (P1) -- (P4) -- (P5) -- cycle;
    \draw[thick, color=blue] (O) -- (P1) -- (P2) -- (P3) -- cycle;
    \draw[thick, dashed] (P5) -- (P8);
    \draw[thick, dashed] (P3) -- (P9);
    \draw[thick, dashed] (P4) -- (P10);
  
    \draw[->, dashed, thick, blue] (P6) -- (P7);
    			
    \node at (1.2, 0.55){(I)};
    \node at (4, 1.3){$G$};
    
    \fill[blue] (O) circle (1.5pt);
    \fill[blue] (P1) circle (1.5pt);
    \fill[blue] (P2) circle (1.5pt);
    \fill[blue] (P3) circle (1.5pt);
    \fill[blue] (P4) circle (1.5pt);
    \fill[blue] (P5) circle (1.5pt);
    \fill[red] (P11) circle (1.5pt);
    \fill[red] (P12) circle (1.5pt);
    \fill[red] (P13) circle (1.5pt);

\end{tikzpicture}
\begin{tikzpicture}
    \draw[->] (0,0)--(6.2,0);
    \draw[->] (0,0)--(0,5);
	\draw (0,0) -- (0,0.1);
	\draw (1.3,0) -- (1.3,0.1);
	\draw (4.7,0) -- (4.7,0.1);
	\draw (6,0) -- (6,0.1);
	\draw (0,0) -- (0.1,0);
	\draw (0,1) -- (0.1,1);
	\draw (0,2) -- (0.1,2);
	\draw (0,2.5) -- (0.1,2.5);
	\draw (0, 4.7) -- (0.1,4.7);
	
    \node[below] at (-0.2,0){0};
    
    \node[below] at (0.55, -0.12){$\dots$};
    \node[below] at (1.3, 0){$t_{m_2}$};
    \node[below] at (3.1, -0.12){$\dots$};
    \node[below] at (4.7, 0){$t_{N}$};
    \node[below] at (5.9, 0){$t_{N+{m_2}-1}$};
    \node[right] at (6.3, 0){$t$};
    	
    \node[left] at (-0.12, 0.5){$\vdots$};		
    \node[left] at (0, 1){$t_{m_1}$};
    \node[left] at (0, 2){$t_j$};
    \node[left] at (0, 2.5){$t_{j+n-1}$};
    \node[left] at (0, 4.7){$t_{N}$};
    \node[left] at (0, 5.1){$t'$};
    \node[left] at (0.25, 5.6){\textbf{Step 2:}};
    			
    \coordinate (O) at (0, 0);
    \coordinate (P1) at (1.3, 0);
    \coordinate (P2) at (2.3, 1);
    \coordinate (P3) at (1, 1);
    \coordinate (P4) at (6, 4.7);
    \coordinate (P5) at (4.7, 4.7);
    \coordinate (P8) at (0, 4.7);
    \coordinate (P9) at (0, 1);
    \coordinate (P10) at (6, 0);
    
    \draw[thick, color=blue] (O) -- (P1) -- (P2) -- (P3) -- cycle;
    \draw[thick, dashed, blue] (O) -- (P1) -- (P4) -- (P5) -- cycle;
    \draw[thick, dashed] (P5) -- (P8);
    \draw[thick, dashed] (P3) -- (P9);
    \draw[thick, dashed] (P4) -- (P10);
    
    \coordinate (P6) at (2, 2);
    \coordinate (P7) at (3.3, 2);
    \coordinate (P11) at (2.5, 2.5);
    \coordinate (P12) at (3.8, 2.5);
    \coordinate (P13) at (6, 2.5);
    \coordinate (P14) at (5.5, 2);
    \coordinate (P15) at (4, 2.25);
    \coordinate (P16) at (5.55, 2.25);
    \coordinate (P17) at (0, 2);
    \coordinate (P18) at (0, 2.5);
    
    \draw[thick, color=red] (P6) -- (P7) -- (P12) -- (P11) -- cycle;
    \draw[thick, dashed, color=red] (P7) -- (P12) -- (P13) -- (P14) -- cycle;
    \draw[thick, dashed] (P6) -- (P17);
    \draw[thick, dashed] (P11) -- (P18);
    
    \draw[->, dashed, thick, red] (P15) -- (P16);
    			
    \node at (1.2, 0.55){(I)};
    \node at (3, 2.25){(II)};
    \node at (4, 1.3){$G$};
    
    \fill[blue] (O) circle (1.5pt);
    \fill[blue] (P1) circle (1.5pt);
    \fill[blue] (P2) circle (1.5pt);
    \fill[blue] (P3) circle (1.5pt);
    \fill[blue] (P4) circle (1.5pt);
    \fill[blue] (P5) circle (1.5pt);
    \fill[red] (P6) circle (1.5pt);
    \fill[red] (P7) circle (1.5pt);
    \fill[red] (P11) circle (1.5pt);
    \fill[red] (P12) circle (1.5pt);
    \fill[red] (P13) circle (1.5pt);
    \fill[red] (P14) circle (1.5pt);
    		
\end{tikzpicture}
\caption{Illustration of the two-step DMD for the Green's function $G(t, t')$ at a fixed $k$-point.}
\label{fig:method}
\end{figure}

An alternative scheme, which only requires taking snapshots within the lower triangular time window $t' \leq t$ for a fixed $k$-point is to sample along the diagonal and subdiagonals of $G(t,t')$, i.e., we can construct the snapshot matrix as 
\begin{equation}
    \mathbf{X} = 
    \begin{bmatrix}
     G(t_1,t_1)   & G(t_2,t_2)     & \cdots & G(t_m,t_m)       \\
     G(t_2,t_1) & G(t_3,t_2)     & \cdots & G(t_{m+1},t_m) \\
     \vdots   & \vdots         & \cdots &  \vdots        \\
     G(t_m,t_1) & G(t_{m+1},t_2) & \cdots & G(t_{2m-1},t_m)
    \end{bmatrix}.
    \label{eq:Xsampdiag}
\end{equation}
The matrix elements contained in \eqref{eq:Xsampdiag} correspond to values of $G(t,t')$ evaluated within the parallelogram outlined in blue and marked as area (I) in Figure~\ref{fig:method}.  Each column of $\mathbf{X}$ corresponds to each row of the $G(t,t')$ within that parallelogram.
In general, the number of rows in $\mathbf{X}$ can be different from the number of columns, i.e., we can sample along the $t'$ direction up to $t' = t_{m_1} = (m_1-1)\Delta t$, and along the $t$ direction up to $t = t_{m_2}$ for $t'=0$ and $t = t_{m_1+m_2-1}$ for $t' = t_{m_1}$ (starting from $t=t_{m_1}$) as shown in Figure~\ref{fig:method}.

Once we perform a DMD or HODMD on this snapshot matrix, we can extrapolate along the diagonal and subdiagonals of $G(t,t')$ first, as indicated by the dashed arrow in the left panel of Figure~\ref{fig:method}.  

To predict values of $G(t,t')$ outside of the subdiagonal bands in the second step, we can sample within a parallelgram time window outlined in red and marked by area (II) in the right panel of Figure~\ref{fig:method} to construct a snapshot matrix of the form
\begin{equation}
        \mathbf{X} = 
    \begin{bmatrix}
     G(t_j,t_j)         & G(t_{j+1},t_j)     & \cdots & G(t_{j+m_2-1},t_j)       \\
     G(t_{j+1},t_{j+1}) & G(t_{j+2},t_{j+1})     & \cdots & G(t_{j+m_2},t_{j+1}) \\
     \vdots   & \vdots         & \cdots &  \vdots        \\
     G(t_{j+n-1},t_{j+n-1}) & G(t_{j+n},t_{j+n-1}) & \cdots & G(t_{j+m_2+n-2},t_{j+n-1})
    \end{bmatrix},
    \label{eq:Xsampoffdiag}
\end{equation}
where $n$ denotes the number of rows in $t'$ we consider together. Note that the sampling window (I) is a special case of (II). It can be used in a DMD analysis to predict values of $G(t,t')$ to the right of the parallelgram region in the second step. However, the snapshot matrix used in this step is the transpose of the $\mathbf{X}$ matrix defined in \eqref{eq:Xsampdiag}.

As in this method, the $k$-point is always fixed, we call it fixed $k$-point (FK) DMD extrapolation.

\section{Results and discussions} \label{sec:example}
In this section, we demonstrate and compare methods for predicting the off-diagonal elements of $G(t,t') $ discussed in the previous section. All methods are applied to 
the two-band Hubbard model problem~\cite{hubbard1963electron,tasaki1998hubbard,white1989numerical} in which the many-body Hamiltonian is given by 
\begin{equation}\label{eq:Htotal}
    H_{\rm{total}}(t) = H_{\rm{s}} + H_{\rm{ext}}(t).
\end{equation}
In \eqref{eq:Htotal}, $H_{\rm{s}}$ is the system Hamiltonian defined as
\begin{align}
    H_{\rm{s}}=\sum_\vk(\epsilon_{v\vk}c^\dagger_{v\vk} c_{v\vk} + \epsilon_{c\vk}c^\dagger_{c\vk}c_{c\vk})-U\sum_k c^\dagger_{c\vk}c_{c\vk}+\frac{U}{N_0}\sum_{\vk_1,\vk_2,\vq}c^\dagger_{v\vk_1+\vq}c^\dagger_{c\vk_2-\vq}c_{c\vk_2}c_{v\vk_1},
    \label{eq:ham}
\end{align}
where $\epsilon_{v\vk}$ ($\epsilon_{c\vk}$) is the band energy of the valence (conduction) band with momentum $\vk$, $U$ is the on-site interaction between the two bands, and $N_0$ is the number of sites in the system. The energy dispersion is given by \begin{align*}
    \epsilon_{v\vk}&=-2(1-\cos(\vk)) - E_g/2 \\
    \epsilon_{c\vk}&=2(1-\cos(\vk)) + E_g/2,
\end{align*}
with $E_g=1$ as the band gap.
The second term in \eqref{eq:Htotal} is the light-matter coupling within the dipole approximation defined by
\begin{align}
    H_{\rm{ext}}(t) = E(t)\sum_{\vk}(d_\vk c^\dagger_{c\vk}c_{v\vk}+d^*_\vk c^\dagger_{v\vk}c_{c\vk}),
    \label{eq:dipole}
\end{align}
where $E(t)$ is the time-dependent intensity of the field, and $d_\vk$ is the dipole matrix element. For simplicity we set $d_\vk=1$.

The many-body Hamiltonian \eqref{eq:Htotal} describes how electrons and holes interact with each other and with a classical light field. 

We seek to solve the KBE associated with the non-equilibrium many-body dynamics associated with this Hamiltonian. The single-particle Hamiltonian $H(t)$ in \eqref{eq:kbe} is derived from the many-body perturbation theory. The self-energy term $\Sigma(t, \overline{t})$ is approximated by the second-Born correction
\begin{align*}
    \Sigma^{\rm{2B}}_{jm}(\mathbf{k}; t,t')=&\frac{U^2}{N^2}\sum_{\mathbf{q}\mathbf{k'}}G_{ps}(\mathbf{k'+q}; t,t')G_{sp}(\mathbf{k'}; t',t)G_{jm}(\mathbf{k-q}; t,t')\\
    & \; -\frac{U^2}{N^2}\sum_{\mathbf{q}\mathbf{k'}}G_{jp}(\mathbf{k'}; t,t')G_{ps}(\mathbf{k'-q}; t',t)G_{sm}(\mathbf{k-q}; t,t'),
\end{align*}
where $G_{pq}(\mathbf{k}; t,t')$ is the two-time Green's function with band indices $p$, $q$, and crystal momentum index $\mathbf{k}$.
As discussed in Section \ref{sec:intro}, the KBEs can be solved numerically by a Runge-Kutta type of time integrator in two times~\cite{kennedy2016diagonally}. Before using the DMD method to predict the long time behavior of the two-time Green's function $G(t,t')$, we first solve the KBEs by numerical time evolution within a relatively small two-time window, and use the numerical solution to  construct data matrices required to perform a DMD.

In the following numerical examples, we assume that $E(t)$ is an instantaneous pulse given by $I\delta(t)$, where $I$ denotes the pulse intensity. The Brillouin zone $[-\pi,\pi]$ is discretized uniformly by $n_k=20$ $k$-points set to $k_s = -\pi + 2(s-1)\pi/n_k$ $(s=1, ..., n_k)$. Our goal is to predict the values of $G(k_s;t,t')$ for $(t,t') \in [0,200] \times [0,200]$ on a uniform two-time grid $(t_i,t_j)$, with $t_i=(i-1)\Delta t$ and $t_j=(j-1)\Delta t$  where $\Delta t = 0.1$, $i$, $j=1, ..., 2001$. As a result, the number of time grid points in each time direction is $N=2001$, and the total number of $G(k_s, t_i,t_j)$'s to be evaluated is $k_s \times N \times N = 20\times 2001 \times 2001$.

In all the numerical experiements presented below, we use HODMD instead of the standard DMD in order to compensate for the potential lack of momentum resolution in the sampled snapshots. Although we could improve the momentum resolution by generating more $k$-points, this approach would significantly increase computational and memory cost used to solve the KBE numerically (within the same two-time window.)~\cite{HODMD}. Following the notation established in Section~\ref{sec:dmd}, we use HODMD($d$) to denote the version of HODMD in which $d$ consecutive snapshots of $G$ (in some time direction) are combined into a single column of the snapshot matrix, and no overlap exists between two adjacent columns of the snapshot matrix, as defined in \eqref{eq:DMDc-d1}.

\subsection{Predicting $G(k; t,t')$ for fixed $t-t'$}
We first report the effectiveness of using HODMD to predict $G(k; t,t')$ for fixed $t-t'$ values, i.e., we predict the values of $G(k;t,t')$ along the time diagonal and subdiagonals within the parallelgram outlined by the bluedashed lines in Figure~\ref{fig:subwindow}. To predict the values of $G(k;t,t')$ for $t-t'=(j-1)\Delta t$, with $j \in \{1,2,...,m_2\}$, we use $G(k;t_i,t_{i-j+1})$ with $j\leq i\leq j+m_1-1$ to construct a snapshot matrix required in a HODMD calculation. Here the parameter $m_2$ is the total number of time subdiagonals of $G(k;t,t')$ we will predict, and $m_1$ is the minimum number of snapshots we will use to perform the HODMD calculation. The HODMD calculation for each $j$ is independent from others, i.e., the HODMD calculations for different time subdiagonals of $G$ can be performed in parallel.  In order to perform these HODMD calculations, we need to first solve the KBE numerically within the time window $[0,t_{m_1+m_2-1}] \times [0,t_{m_1+m_2-1}]$.  When $t-t' < (m_2-1) \Delta t$, more snapshots can be used in the HODMD analysis. In particular, for $t=t'$, we can use as many as $m_1+m_2-1$ snapshots. 

\begin{figure}[t!]
	\qquad
	\begin{subfigure}{0.42\textwidth}
		\centering
		\includegraphics[width=1\textwidth]{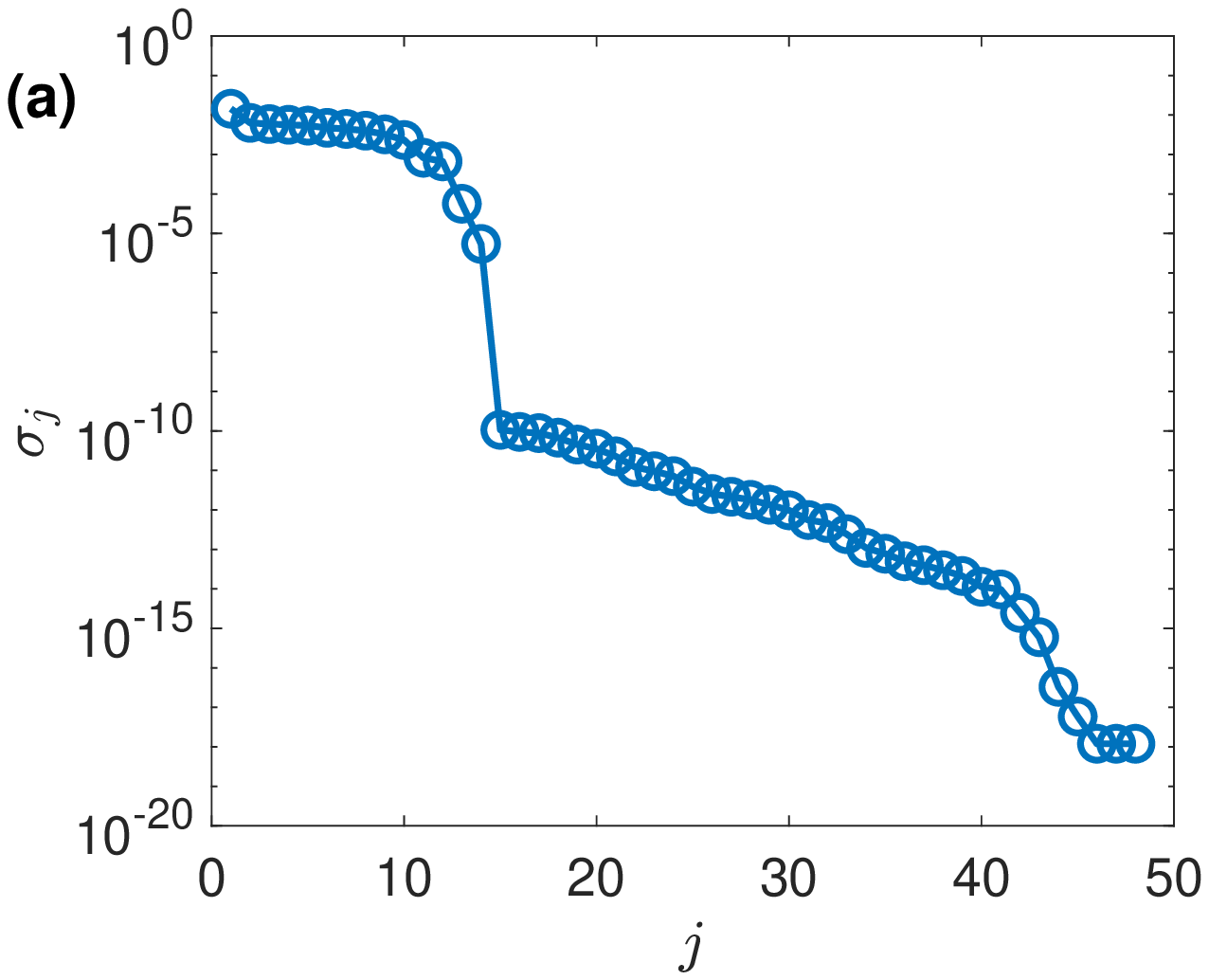}
	\end{subfigure}%
	\begin{subfigure}{0.42\textwidth}
		\centering
		\includegraphics[width=1\textwidth]{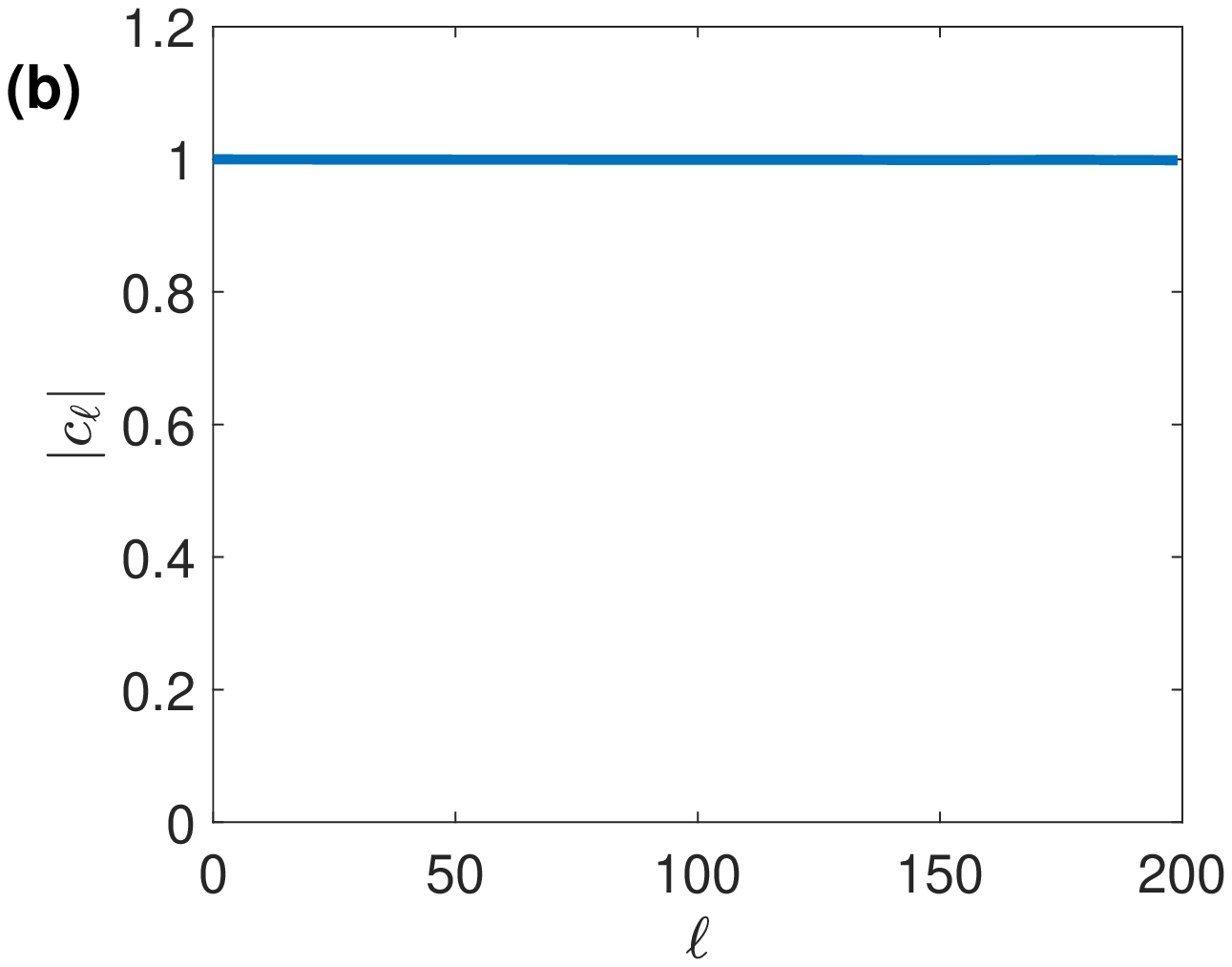}
	\end{subfigure}
	\caption{$I=0.001$. \textbf{(a)} The singular values of the snapshot matrix $\widetilde{\mathbf{X}}_1$ \eqref{eq:DMDc-d1} in HODMD(4) constructed from $G(k; (m+100)\Delta t, m\Delta t)$, for $m=0,1,...,m_1-1=99$; \textbf{(b)} The correlation $|c_{\ell}|$ between the numerical solution of the KBE and the HODMD(4) extrapolation of $G$ along $(t+\ell \Delta t,t)$, $\ell=0,1,...,m_2-1=199$.}
	\label{fig:Amp0p001_sc1}
\end{figure}

Figure~\ref{fig:Amp0p001_sc1}(a) shows the singular values of the snapshot matrix $\widetilde{\mathbf{X}}_1$ \eqref{eq:DMDc-d1} constructed for $t-t'=100\Delta t=10$.  The intensity of the external pulse in \eqref{eq:dipole} is set to $I=0.001$. We set $m_1$ to $100$ to include at least $100$ snapshots in the snapshot matrix $\mathbf{X}$, and use HODMD($4$) to perform the extrapolation. In this case, the singular values of $\mathbf{X}$ decay rapidly. Only the leading 14 singular values are significantly larger than 0, indicating that the dynamics of the $G(k;t+100\Delta t,t)$ can be well characterized by 14 DMD modes. 

To assess the accuracy of the HODMD extrapolation, in Figure~\ref{fig:Amp0p001_sc1}(b), we plot the correlation  $|c_\ell|$ between the numerical solution of the KBE and the HODMD extrapolation along $(t+\ell \Delta t, t)$, which is defined by
\begin{equation}\label{eq:err1}
	c_{\ell} = \min_{s}\frac{\langle G(k_s; t+\ell\Delta t, t),G^{\mathrm{DMD}}(k_s; t+\ell\Delta t, t)\rangle}{\|G(k_s; t+\ell\Delta t, t) \| \|G^{\mathrm{DMD}}(k_s; t+\ell\Delta t, t)\|}, \quad \ell = 0, 1, ..., 199,
\end{equation}
where $G(k_s; t+\ell\Delta t, t)$ is obtained from the numerical solution of the KBE on a uniform two-time grid in $[0,200] \times [0,200]$, $G^{\rm{DMD}}(k_s; t+\ell\Delta t, t)$ is the extrapolated trajectory produced from HODMD, and $\langle \cdot, \cdot \rangle$ denotes the standard Euclidean inner product of two complex vectors. We note that $c_{\ell}$ is the cosine of the angle between the predicted and the computed trajectories. For each $\ell$, we take the minimum of such cosine values among all $k$-points, which yields the largest difference between the extrapolated and the original trajectories among all $k$ points. We can clearly see that for $m_1=100$ and $m_2=200$, the HODMD prediction is nearly perfect for $I=0.001$.

\begin{figure}[t!]
	\centering
	\includegraphics[width=0.9\textwidth]{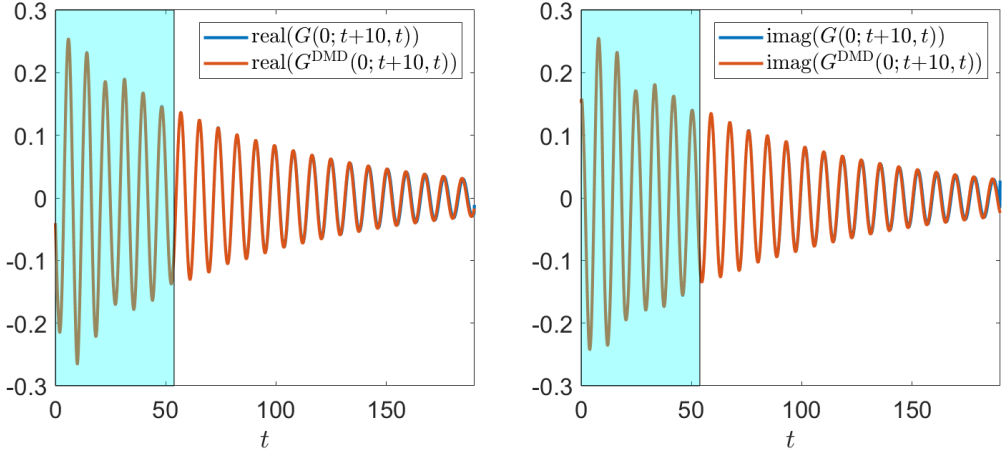}
	\caption{$I=0.5$. The extrapolated $G(0; t+10, t)$ by HODMD(5) where the snapshot matrix $\widetilde{\mathbf{X}}_1$ \eqref{eq:DMDc-d1} is constructed from $G(k; (m+100)\Delta t, m\Delta t)$, for $m=0,1,...,m_1-1=449$. The shaded region represents the window of sampled snapshots from numerical solution of the KBE.}
	\label{fig:Amp0p5_diag1}
\end{figure}

When $I$ is increased to 0.5, we perform a HODMD(5) extrapolation using $m_1=450$, which is the minimum number of snapshots required to produce satisfactory extrapolations along the diagonals, and $m_2=400$, which is the minimum number of snapshots required for each $t'$ to produce a satisfactory extrapolation away from the diagonal (See section~\ref{sec:method}). 
The singular values of the snapshot matrix for $G(k; t+100\Delta t, t)$ and the correlation $|c_{\ell}|$ between the numerical solution of the KBE and the HODMD prediction in are similar to those shown in Figure~\ref{fig:Amp0p001_sc1} where $I=0.001$. The only difference is that the snapshot matrix has more large singular values, indicating that the dynamics associated with $I=0.5$ contains more momentum and temporal features than the dynamics associated with a smaller $I$. These additional features would need to be accounted for by an approximate Koopman operator of a larger dimension and thus more terms in \eqref{eq:dmdmodes}. With these terms, all the $m_2=400$ subdiagonals are accurately extrapolated as the values of $|c_\ell|$ are close to $1$ for all $\ell$. The extrapolated trajectory of the $100$-th subdiagonal of $G$, i.e., $G^{\rm{DMD}}(0; t+10, t)$ is plotted in Figure~\ref{fig:Amp0p5_diag1} and compared with the trajectory $G(0; t+10, t)$ obtained from the numerical solution of the KBE. The sampled data are marked by the blue shaded window. We can observe that the extrapolated trajectory successfully captures the oscillating frequency and the decay rate of the amplitude.


\begin{figure}[t!]
	\qquad
	\includegraphics[width=0.42\textwidth]{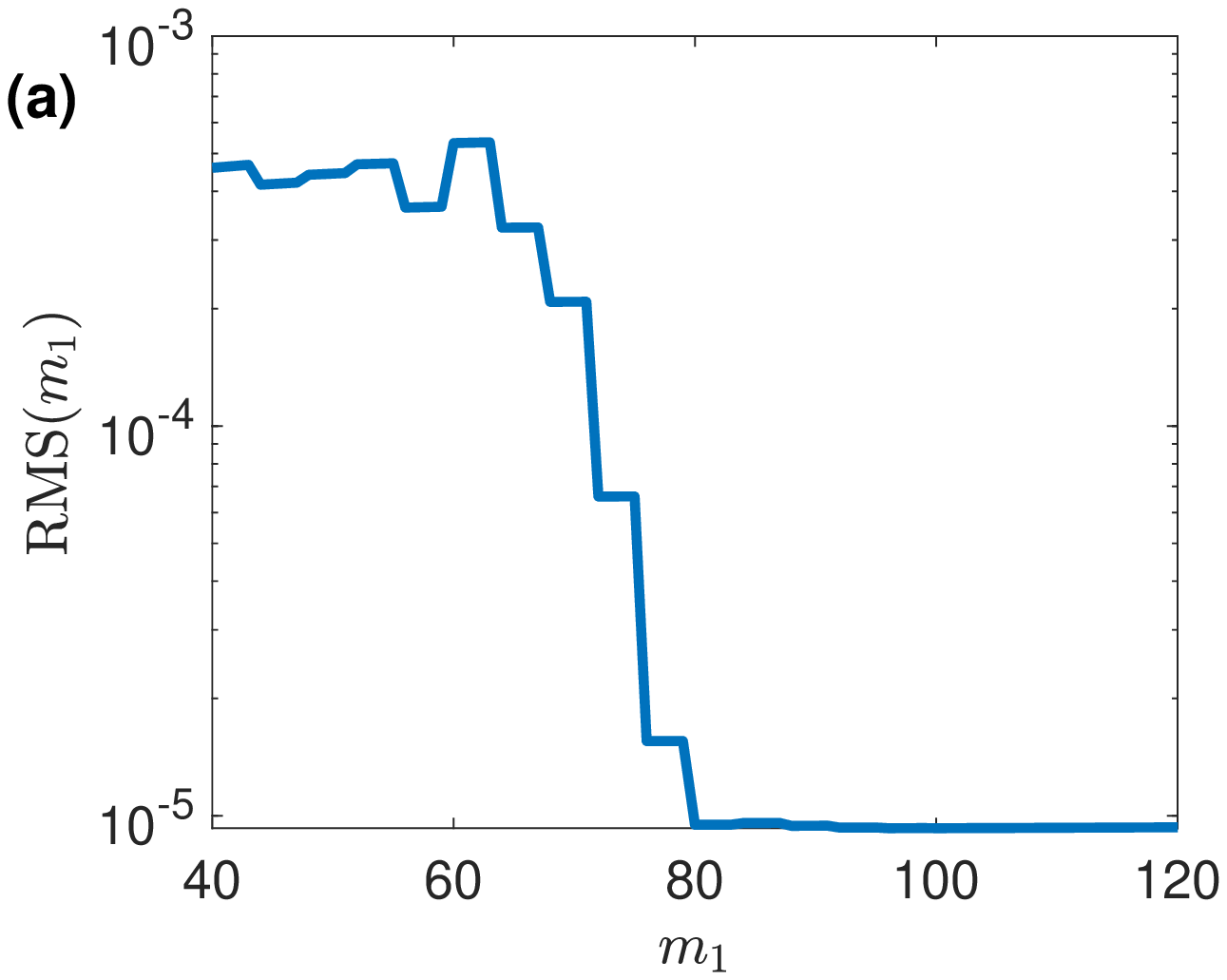}
	\includegraphics[width=0.42\textwidth]{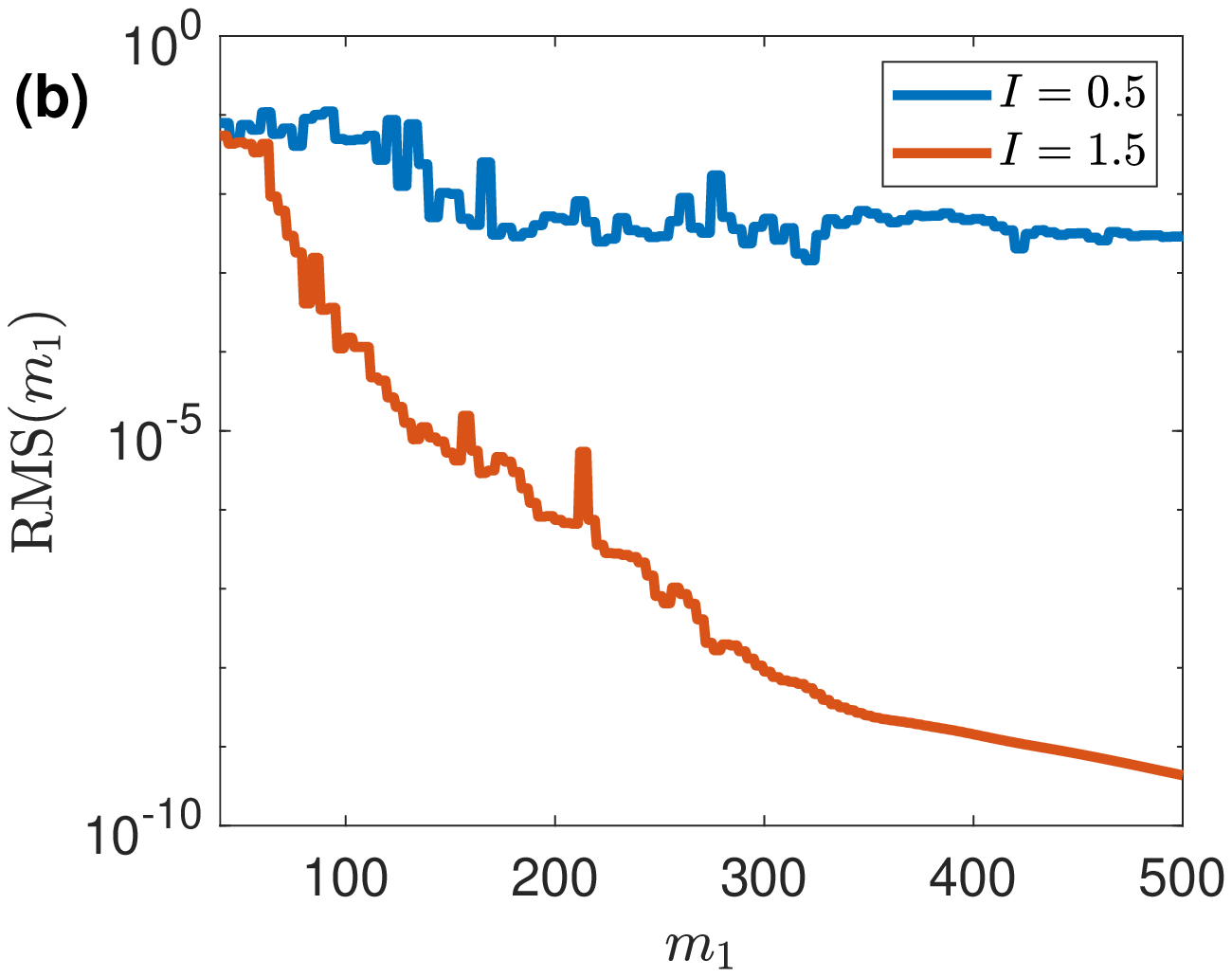}
	\caption{The root mean square errors ${\rm{RMS}}(m_1)$ for $G(k; t+10, t)$ when \textbf{(a)} $I=0.001$; \textbf{(b)} $I=0.5$ and $I=1.5$. }
	\label{fig:err1}
\end{figure}

To check the number of snapshots ($m_1$) required in the HODMD to accurately extrapolate $G(t+10,t)$, we define the root mean square (RMS) error of the extrapolated trajectory as
\begin{equation}
	{\rm{RMS}}(m_1) = \left[\frac{\sum_{j=1}^{n_k}\sum_{p=m_1+1}^{N-m_2+1}|G(k_j; t_{p+100}, t_{p}) - G^{\rm{DMD}}(k_j; t_{p+100}, t_{p})|^2}{n_k(N-m_1-m_2+1)}\right]^{1/2},
\end{equation}
where we set the value of $m_2$ to 101.
The extrapolation is computed by HODMD($4$) when $I=0.001$, HODMD($5$) when $I=0.5$ and HODMD($4$) when $I=1.5$. The corresponding RMS errors are plotted in Figure~\ref{fig:err1}. 

As we can see from Figure~\ref{fig:err1}(a), the RMS for the extrapolated $G(t+10,t)$ starts to decrease rapidly when $m_1 > 60$ and levels off around $m_1=80$ in the $I=0.001$ case. The magnitude of RMS matches that of the numerical integration error contained in the numerical solution of the KBE. 
When $I=0.5$, the RMS starts to decrease from $10^{-1}$ to $10^{-2}$, which is the expected level of error in the numerical solution of KBE when $m_1>120$.  When $I=1.5$, the RMS starts to decrease rapidly when $m_1 > 50$, when $m_1=500$, the RMS is on the order of $10^{-10}$. This is due to the fact that the magnitude of $G(t+10,t)$ rapidly decreases towards 0 when $t$ increases.

\subsection{Predicting $G(k; t,t')$ for fixed $t'$}\label{sec:fixtprime}
As discussed in Section~\ref{sec:method}, to predict the values of the two-time Green's function outside of the $m_2$ subdiagonal bands in $(t,t')$, we can use HODMD to extrapolate $G(k; t, t')$ horizontally on the two-time grid by fixing $t'$ in $G$. For each fixed $t'=t_j$, we use either the computed or extrapolated values of $G(k;t_i,t_j)$, for $j \leq i \leq j+m_2-1$ to construct a snapshot matrix from which DMD modes can be extracted to extrapolate the values of $G(k;t_i,t_j)$ for $j+m_2 \leq i \leq N$.

The accuracy of the prediction can be assessed by examining the correlation between the extrapolated Green's function, denoted by $G^{\rm{DMD}}(k;t,30)$ and the numerical solution of the KBE denoted by $G(k;t,t')$ along a fixed $t'$ defined as
\begin{equation}\label{eq:corr2}
c^k(t') = \frac{\langle G(k; t, t'),G^{\mathrm{DMD}}(k; t, t')\rangle}{\|G(k; t, t') \| \|G^{\mathrm{DMD}}(k; t, t')\|}, \quad k = k_s, \quad s=1, ..., n_k.
\end{equation}

\begin{figure}[t!]
	\centering
	\includegraphics[width=0.9\textwidth]{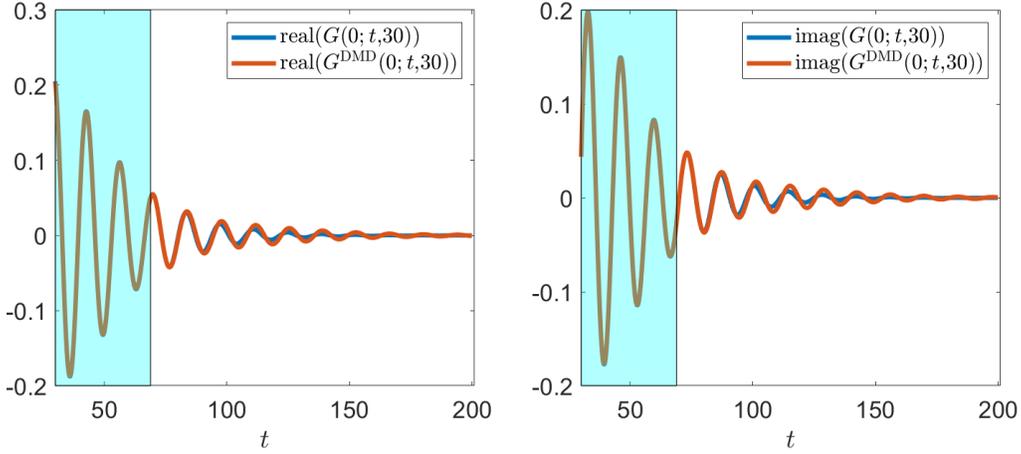}
	\caption{$I=0.5$. A comparision between the real and imaginary parts of $G^{\rm{DMD}}(0; t, 30)$ by HODMD($10$) with those of $G(0;t,30)$. The shaded area marks time window from which snapshots are used to construct the HODMD model.}
	\label{fig:Amp0p5_md1_traj1}
\end{figure}

When the intensity of the external field $E(t)=I\delta(t)$ in \eqref{eq:dipole} is set to $I=0.5$, we found $c^k(t')$ are to be close to 1.0 for nearly $t'$ and $k$-points.
This agreement is confirmed in Figure~\ref{fig:Amp0p5_md1_traj1} where we show the real and imaginary parts of $G^{\rm{DMD}}(0;t,30)$ match well with those of $G(0;t,30)$.

\begin{figure}[t!]
	\centering
	\includegraphics[width=0.9\textwidth]{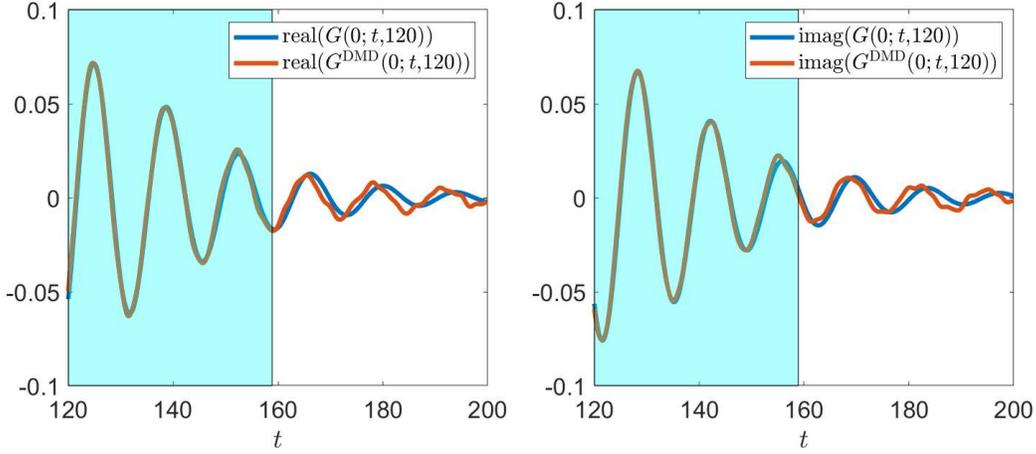}
	\caption{$I=0.5$. A comparision between the real and imaginary parts of  $G^{\rm{DMD}}(0;t,120)$ by HODMD($10$) with those of $G(0;t,120)$. The shaded area marks time window from which snapshots are used to construct the HODMD model.}
	\label{fig:Amp0p5_md1_traj2}
\end{figure}

We make a similar comparison between $G(k;t,120)$ and $G^{\rm{DMD}}(k;t,120)$ in Figure~\ref{fig:Amp0p5_md1_traj2}. Note that, in this case, the snapshot matrix used in HODMD is constructed from the extrapolated values of $G(k,t_j,120)$, $120 \leq t_i \leq 120+m_2-1$, obtained in a previous HODMD step, in which $m_2=400$ subdiagonal lines of $G(;,t,t')$ are extrapolated from the numerical solution of the KBE within $[0,84.9]\times [0,84.9]$.
The correlation factor $|c^k(120)|$ deviates slightly from 1.0 for some $k$ values. Such small deviations can also be seen in Figure~\ref{fig:Amp0p5_md1_traj2} where we plot both the real and imaginary parts of $G(0;t,120)$ and $G^{\rm{DMD}}(0;t,120)$. We believe these  small deviations are caused by small extrapolation errors introduced in the previous step in which HODMD is used to extraploate $G(k;t,t')$ along the diagonal and subdiagonals of the two-time grid.


In Section~\ref{sec:method}, we also discussed the possibility to extrapolate $G(k;t,t')$ horizontally from the numerical solution of the KBE in the time window $[0,t_{m}]\times [0,t_{m}]$ directly for $t'<t_m$ without constructing an HODMD model to extrapolate along the diagonal and subdiagonals of the two-time window first.
We now examine the effectiveness of this approach.

Because the solution to the KBE is computed for $t \geq t'$  within $[0, t_m]\times[0, t_m]$, the larger the $t'$, the fewer data points we can use to construct the snapshot matrix. However, by making use of the following symmetric property of the Green's function, i.e.,
\begin{equation}
	G_{b_1,b_2}(k; t, t') = -\overline{G_{b_2,b_1}(k; t', t)},
\end{equation}
where $b_1$ and $b_2$ denote the band indices, we can augment the snapshot matrix $G_{12}(k;t,t')$ for a fixed $t'$ with samples of $-\overline{G_{b_2,b_1}(k;t',t)}$. Unfortunately, such a symmetry exploiting data augmentation scheme is not always satisfactory as we will see below.

In the following, we set $m$ to 500, i.e., we first solve the KBE within the time window of $[0, t_{500}]\times [0, t_{500}] = [0,49.9]\times [0,49.9]$. We use the values of computed $G_{12}(k;t,10)$ and $G_{12}(k;t,45)$ within 
this time window to construct snapshot matrices that can be used in HODMD(6) to extrapolate $G_{12}(k;t,10)$ and $G_{12}(k;t,45)$ for $t > 49.9$, as shown in Figure~\ref{fig:onedir}.

When $I=0.5$, the singular values of these snapshot matrices decrease rapidly. For example, for both $t'=10$ and $t'=45$, there are around $65$ dominant singular values, and there is a clear gap between these singular values and the others. %

\begin{figure}[t!]
	\qquad
	\includegraphics[width=0.42\textwidth]{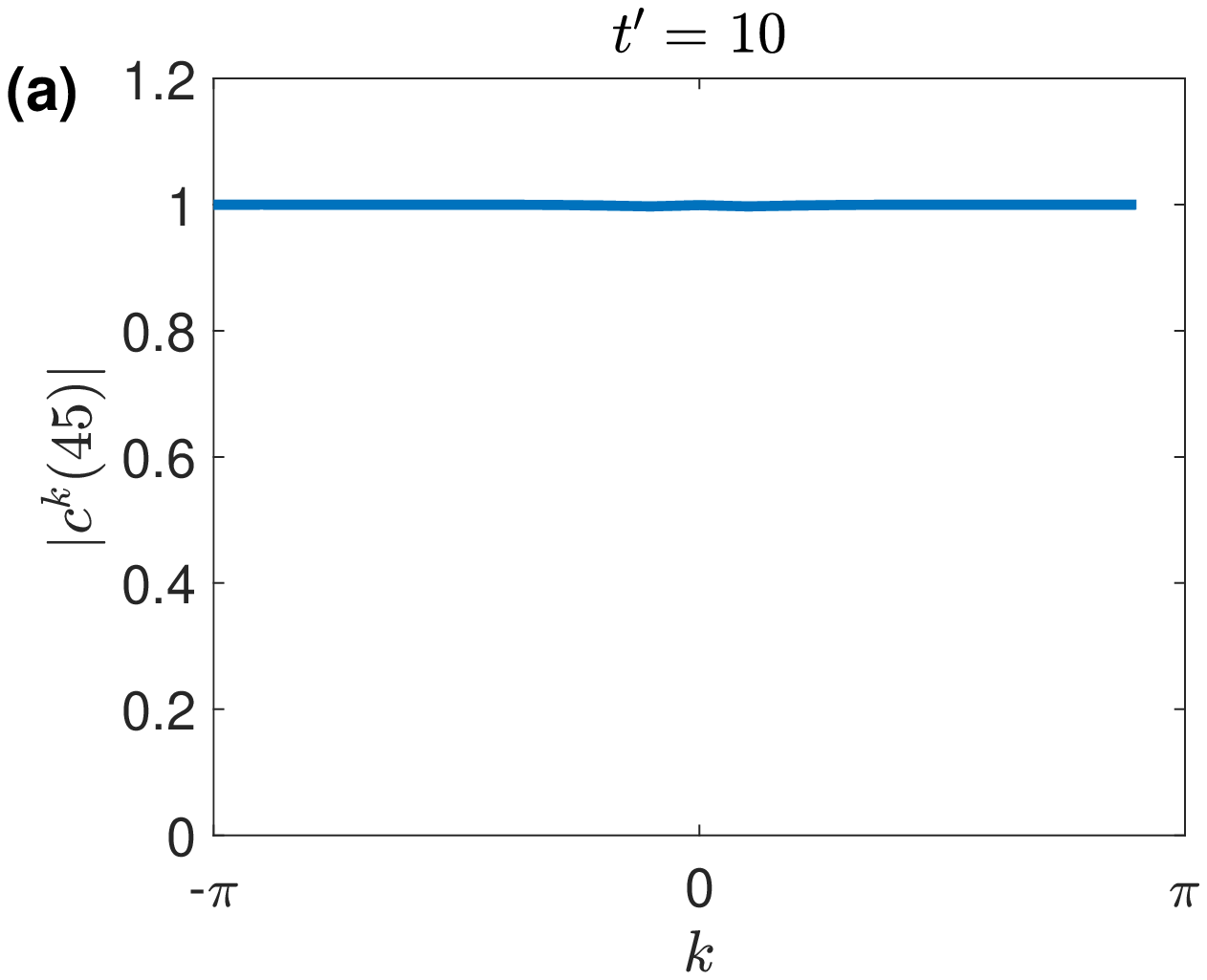}
	\includegraphics[width=0.42\textwidth]{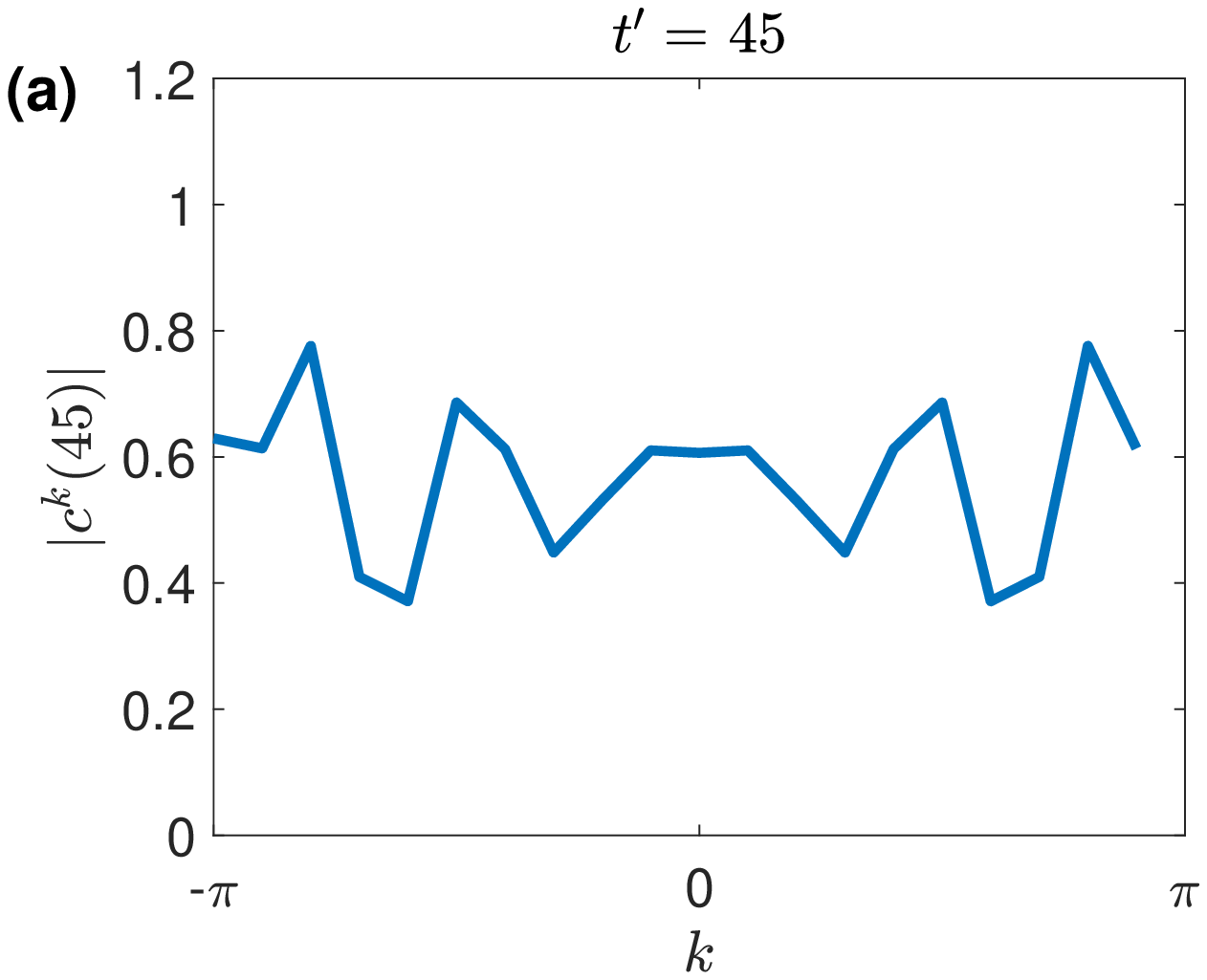}
	\caption{$I=0.5$. \textbf{(a)} The correlation $|c^k(10)|$ between $G^{\rm{DMD}}(k;t,10)$ and $G(k;t,10)$; \textbf{(b)} The correlation $|c^k(45)|$ between $G^{\rm{DMD}}(k;t,45)$ and $G(k;t,45)$.}
	\label{fig:Amp0p5_err}
\end{figure}

However, Figure~\ref{fig:Amp0p5_err}  shows that the two (time) slices of the extrapolated Green's functions exhibit different accuracy features.  At $t'=10$, the correlation between the HODMD extrapolation $G^{\rm{DMD}}(k;t,10)$ and the corresponding $G(k;t,10)$ obtained from the numerical solution of the KBE is nearly perfect at all k points.  Much lower correlation is observed between $G^{\rm{DMD}}(k;t,45)$ and $G(k;t,45)$. 

\begin{figure}[t!]
	\centering
	\includegraphics[width=0.9\textwidth]{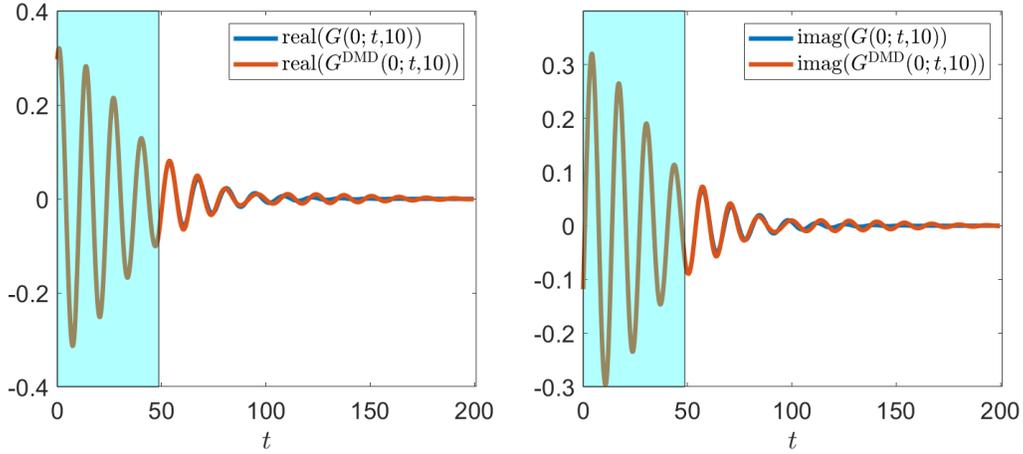}
	\caption{$I=0.5$. The extrapolated $G(0; t, 10)$ by HODMD(6) with the snapshot matrix $\widetilde{\mathbf{X}}_1$ constructed from $G(k; n\Delta t, 10)$, for $n=0, 1, .., m_2-1 = 499$. The shaded area marks time window from which snapshots are used to construct the HODMD model.}
	\label{fig:Amp0p5_md1_G1}
\end{figure}

\begin{figure}[t!]
	\centering
	\includegraphics[width=0.9\textwidth]{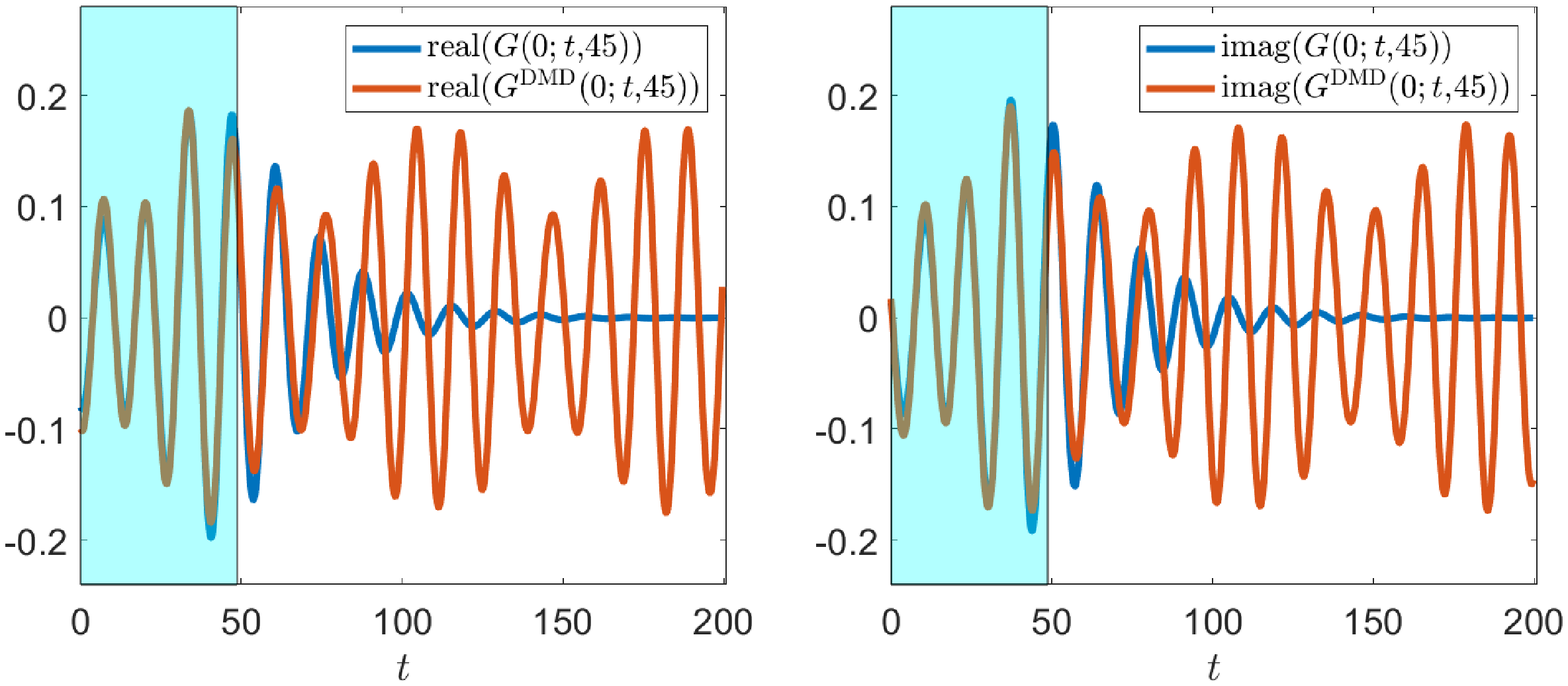}
	\caption{$I=0.5$. The extrapolated $G(0; t, 45)$ by HODMD(6) with the snapshot matrix $\widetilde{\mathbf{X}}_1$ constructed from $G(k; n\Delta t, 45)$, for $n=0, 1, .., m_2-1 = 499$. The shaded area marks time window from which snapshots are used to construct the HODMD model.}
	\label{fig:Amp0p5_md1_G2}
\end{figure}

The excellent agreement between $G^{\rm{DMD}}(k;t,10)$ and $G(k;t,10)$ and the lack of satisfactory agreement between $G^{\rm{DMD}}(k;t,45)$ and $G(k;t,45)$ are confirmed in Figures~\ref{fig:Amp0p5_md1_G1} and~\ref{fig:Amp0p5_md1_G2} where we plot both the real and imaginary parts of the computed and extrapolated $G(0;t,10)$ and $G(0;t,45)$ respectively.
In particular, at $t'=10$, the HODMD extrapolation correctly captures both the decay amplitude and oscillation frequencies of $G(0;t,10)$. However, at $t'=45$, the extrapolated $G^{\rm{DMD}}(0;t,45)$ deviates significantly from the solution of the KBE.

We believe that the reason HODMD performs poorly in predicting the values of 
$G(k;t,45)$ for $t > 49.9$ is that an accurate HODMD model requires a sufficiently
large number of snapshots $G(k;t_i,t')$ for $t_i \geq t'$.
In this experiment, only $50$ snapshots within $[45,50]$ are available for using 
in HODMD, which is apparently not enough to construct an accurate reduced order model.




\subsection{Predicting $G(k;t,t')$ for a fixed $k$}
We now report the effectiveness of an alternative DMD extrapolation scheme discussed in Section~\ref{sec:fixk}. In this scheme, we use snapshots of $G(k;t,t')$ within a small two-time window for a fixed $k$ to construct a DMD-based reduced order model from which values of $G(k;t,t')$ are predicted for larger $t$ and $t'$.

In the first numerical example, we set the intensity of the external field to $I=0.001$ and solve the KBE within the two time window $[0,54.9]\times [0,54.9]$. We then set $m_1 = 250$, $m_2=300$ and construct a snapshot matrix $\mathbf{X}$ for each $k$ according to \eqref{eq:Xsampdiag}. We use HODMD(10) to construct a reduced order model to extrpolate values of $G(k;t,t')$ along the diagonal and subdiagonals of the two time grid for $t-t'\leq 24.9$.

\begin{figure}[t!]
	\centering
	\includegraphics[width=0.9\textwidth]{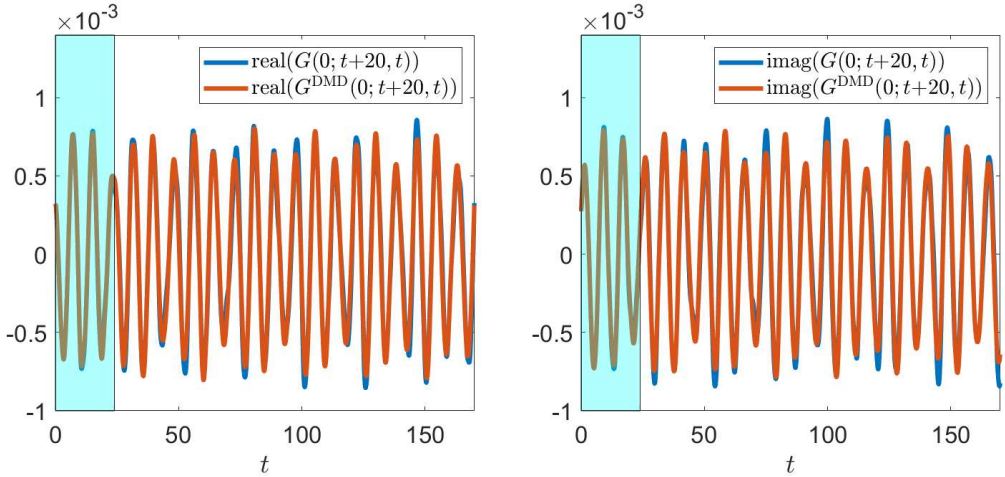}
	\caption{$I=0.001$. A comparison of $G^{\rm{DMD}}(0;t+20,t)$ with $G(0;t+20,t)$. The extrapolation is computed by HODMD(10) with the snapshot matrix $\widetilde{\mathbf{X}}_1$ constructed from $G(0;(p+q)\Delta t , p\Delta t)$ with $p = 0, 1, ..., m_2-1 = 299$ and $q = 0, 1, ..., m_1-1=249$. The shaded area marks time window from which snapshots are used to construct the HODMD model.}
	\label{fig:Amp0p001_md2_traj1}
\end{figure}

Figure~\ref{fig:Amp0p001_md2_traj1} shows that for $k=0$ and $t-t'=20$, the extrapolated $G^{\rm{DMD}}(0,t+20,t)$ agrees well with the computed $G(0;t+20,t)$ obtained from the numerical solution of the KBE. Similar good agreements are observed for other $t-t'$ and $k$ values.

Following the strategy presented in Section~\ref{sec:fixk}, once we extrapolate $G(k;t,t')$ along the diagonal and subdiagonals for each $k$, we then partition the parallelgram formed by the subdiagonal bands on a two time grid vertically into several strips of size $n \times m_2$. In this experiment, we choose $n=30$ and $m_2=300$.   The computed or extrapolated values of $G(k;t,t')$ values within each strip are used to construct a snapshot matrix $\mathbf{X}$ according to \eqref{eq:Xsampoffdiag}. HODMD(2) is performed to construct a reduced order model from which the values of $G(k;t,t')$ outside of the subdiagonal band of width $m_2$ can be extrapolated.

\begin{figure}[t!]
	\centering
	\includegraphics[width=0.9\textwidth]{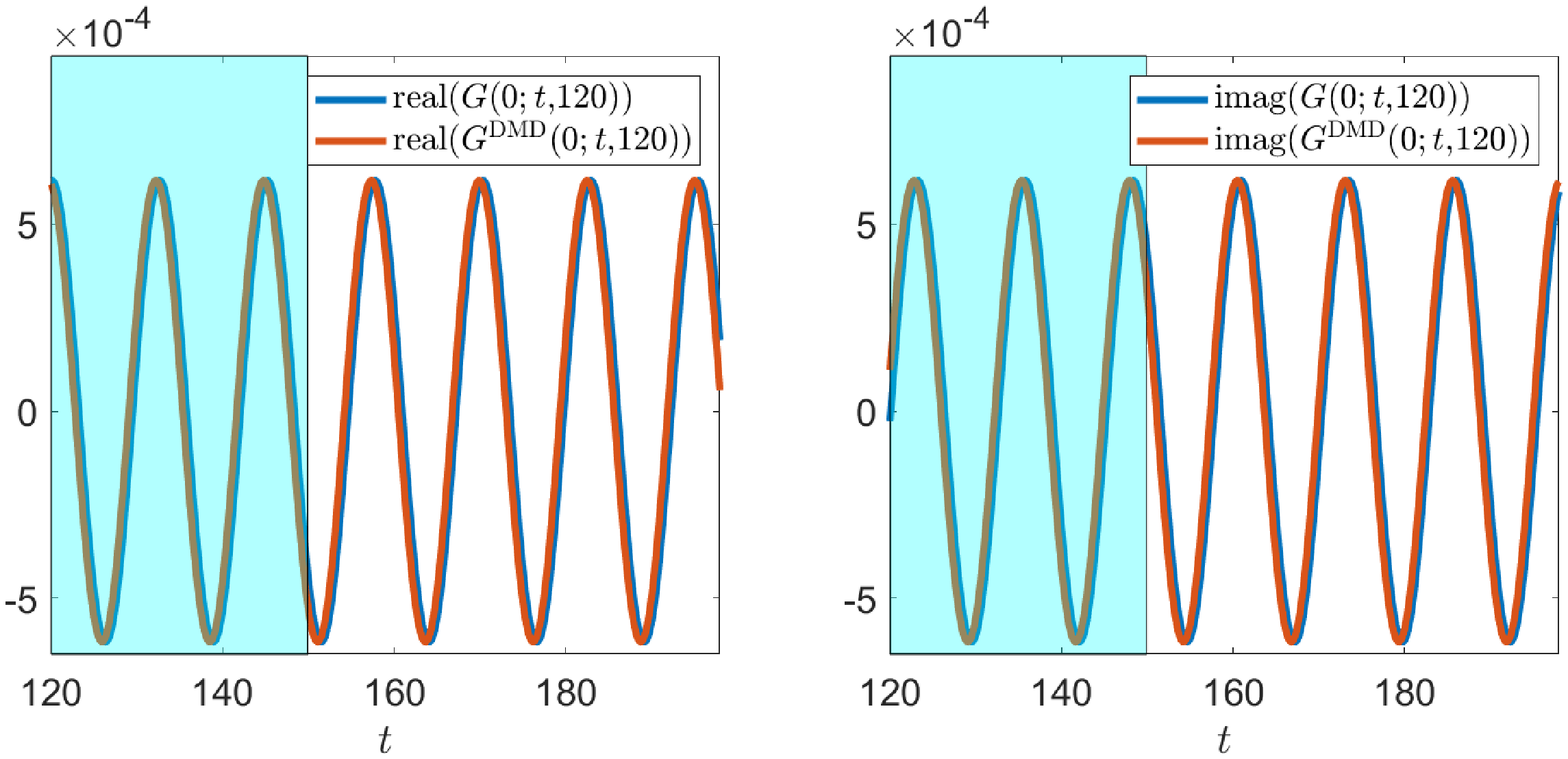}
	\caption{$I=0.001$. A comparison of $G^{\rm{DMD}}(0;t,120)$ with $G(0;t,120)$. The extrapolation is computed by HODMD(2) with the snapshot matrix $\widetilde{\mathbf{X}}_1$ constructed from $G(0;120+(p+q)\Delta t , 120+p\Delta t)$ with $p = 0, 1, ..., n-1 = 29$ and $q = 0, 1, ..., m_2-1=299$. The shaded area marks time window from which snapshots are used to construct the HODMD model.}
	\label{fig:Amp0p001_md2_traj2}
\end{figure}

Figure~\ref{fig:Amp0p001_md2_traj2} shows that, for $k=0$ and $t'=120$, the extrapolated $G^{\rm{DMD}}(0,t,120)$ agrees well with the computed $G(0;t,120)$ obtained from the numerical solution of the KBE. Note that in this case, the snapshot matrix $\mathbf{X}$ is constructed from the extrapolated values of $G(0,t,t')$ along the subdiagonals of the two-time grid in the previous step.  Similar good agreements are observed for other $t'$ and $k$ values.

These $m_1$ and $m_2$ values used in the above experiments appear to be the minimal required to produce accurate extrapolations both along the diagonals and off-diagonals.  The choice of $n=30$ is somewhat arbitrary. We observe that the extrapolation in the $t$ direction is accurate for several values of $n \in [10,1000]$. 

%

\begin{figure}[t!]
	\qquad
	\includegraphics[width=0.42\textwidth]{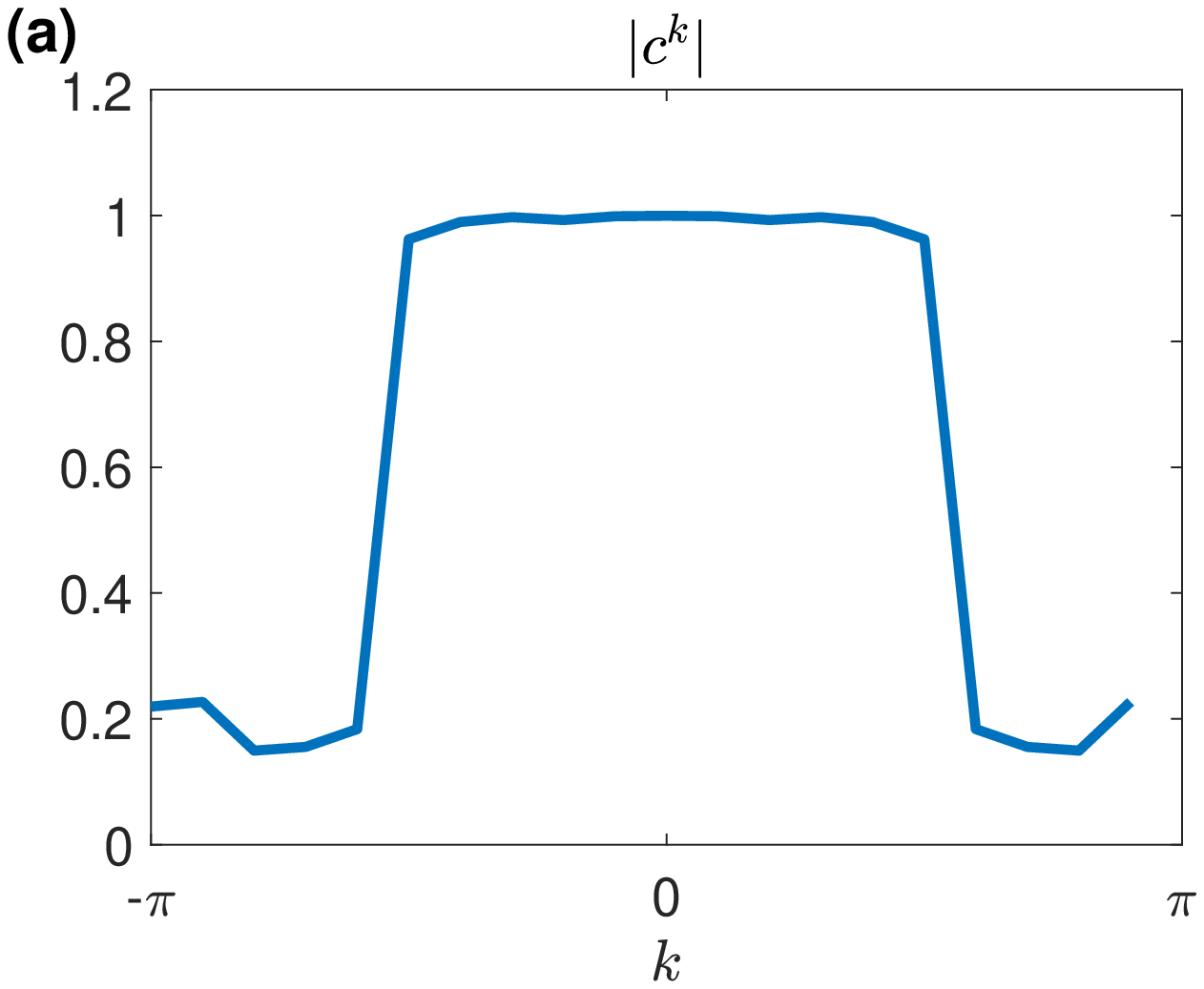}
	\includegraphics[width=0.42\textwidth]{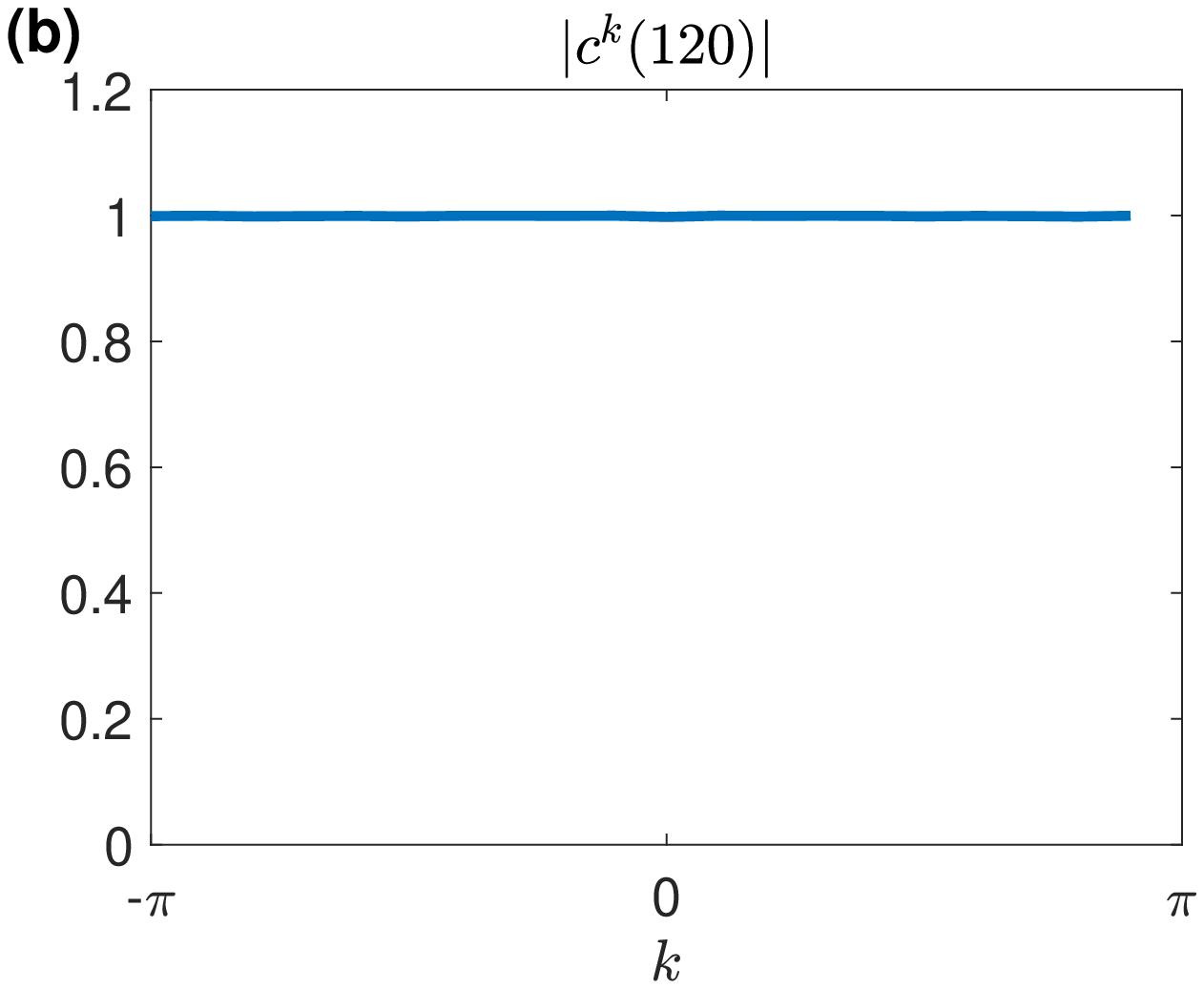}
	\caption{$I=0.5$. \textbf{(a)} The correlation $|c^k|$ between the numerical solution of the KBE and the HODMD(10) extrapolation of $G(k; t+20;t)$. The snapshot matrix $\widetilde{\mathbf{X}}_1$ is constructed from $G(k;(p+q)\Delta t, p\Delta t)$ with $p = 0, 1, ..., m_2-1 = 299$ and $q = 0, 1, ..., m_1-1=449$; \textbf{(b)} The correlation $|c^k(120)|$ between the numerical solution of the KBE and the HODMD(5) extrapolation of $G(k; t, 120)$. The snapshot matrix $\widetilde{\mathbf{X}}_1$ is constructed from $G(k;120+(p+q)\Delta t, 120+p\Delta t)$ with $p = 0, 1, ..., n-1 = 29$ and $q = 0, 1, ..., m_2-1=299$.}
	\label{fig:Amp0p5_md2_err}
\end{figure}

\begin{figure}[t!]
	\centering
	\includegraphics[width=0.9\textwidth]{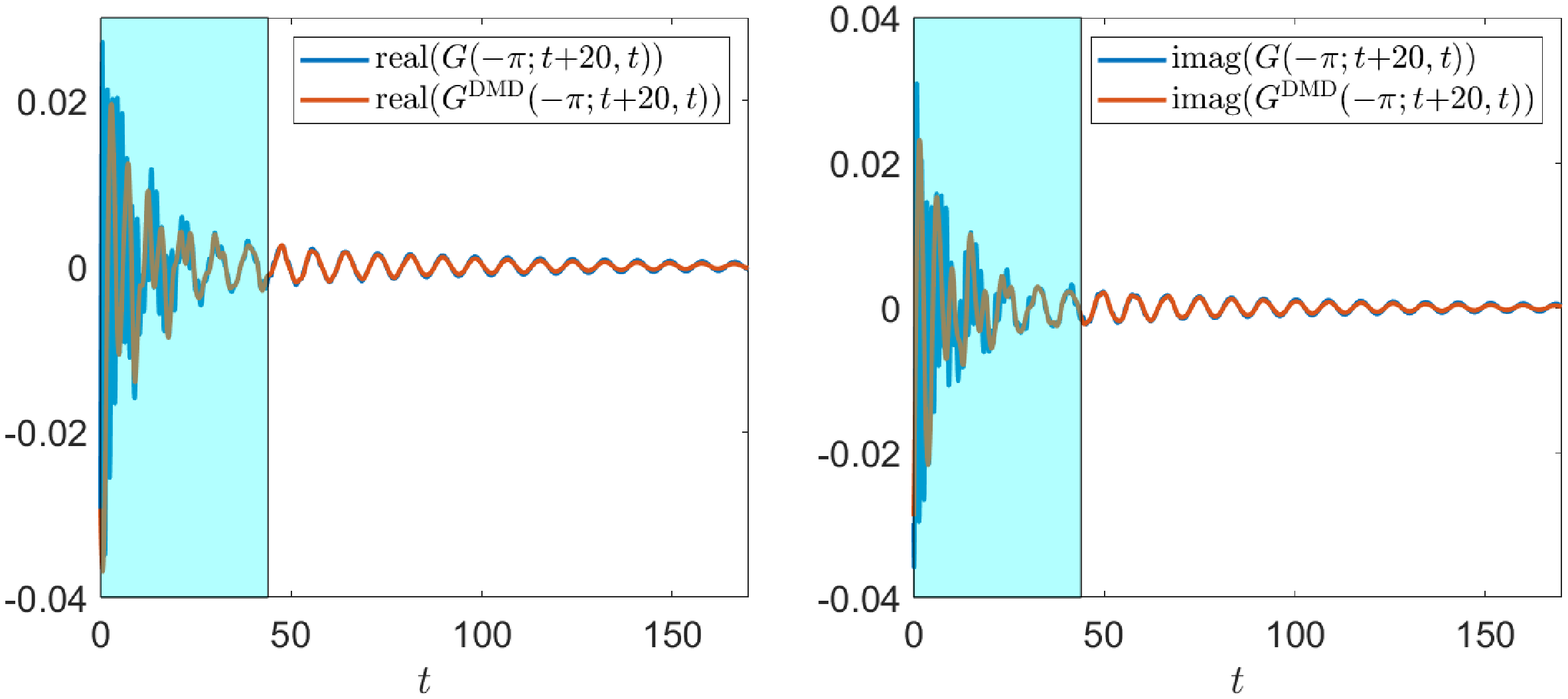}
	\caption{$I=0.5$. A comparison of $G^{\rm{DMD}}(-\pi;t+20,t)$ with $G(-\pi;t+20,t)$. The extrapolation is computed by HODMD(10) with the snapshot matrix $\widetilde{\mathbf{X}}_1$ constructed from $G(-\pi;(p+q)\Delta t, p\Delta t)$ with $p = 0, 1, ..., m_2-1 = 299$ and $q = 0, 1, ..., m_1-1=449$. The shaded area marks time window from which snapshots are used to construct the HODMD model.}
	\label{fig:Amp0p5_md2_step1}
\end{figure}

When we increase the intensity of the external field to $I=0.5$, we need to increase the value of $m_1$ to $m_1=450$ in order to obtain accurate extrapolation of $G(k;t,t')$ along the $m_2=300$ subdiagonals of the two-time grid.  We use HODMD($10$) to construct the reduced order model used to extrapolate $G(k,t,t')$ along the subdiagonals of the two-time grid for each $k$.  The correlation $|c^k|$ between $G^{\rm{DMD}}(k;t+20,t)$ and $G(k;t+20,t)$ (obtained from the numerical solution of the KBE) is shown for all $k$ points in Figure~\ref{fig:Amp0p5_md2_err}(a).  Although the correlations for the first $5$ and the last $4$ $k$-points appear to be somewhat low, the extrapolation appears to be accurate outside the time window that contains the sampled snapshots, as shown  in Figure~\ref{fig:Amp0p5_md2_step1}.  This observation suggests that, for a fixed $k$, the long-time dynamics of $G(k;t,t')$ can be well approximated by a linear reduced order model along a fixed $t-t'$ for $t-t'\leq 29.9$, even though the model does not quite fit $G(k;t,t')$ for small $t$ and $t'$ on which $G(k;t,t')$ behaves more nonlinearly due to the onset of a higher intensity pulse. Since the values of $G(k;t,t')$ are already available from the numerical solution of the KBE for small $t$ and $t'$, and because we are mainly interested in the dynamics of $G$ for large $t$ and $t'$, the extrapolation produced by HODMD(10) in this case is acceptable.

\begin{figure}[t!]
	\centering
	\includegraphics[width=0.9\textwidth]{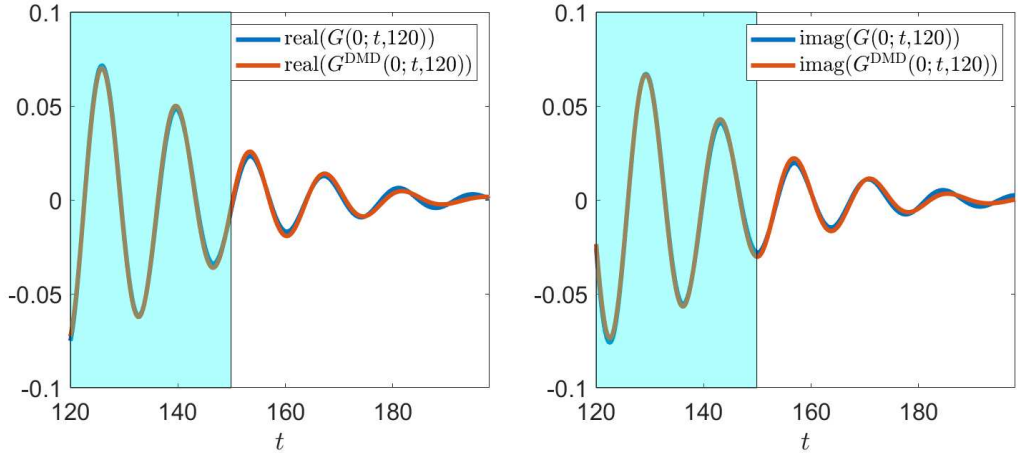}
	\caption{$I=0.5$. A comparison of $G^{\rm{DMD}}(0;t,120)$ with $G(0;t,120)$. The extrapolation is computed by HODMD(5) with the snapshot matrix $\widetilde{\mathbf{X}}_1$ constructed from $G(0;120+(p+q)\Delta t, 120+p\Delta t)$ with $p = 0, 1, ..., n-1 = 29$ and $q = 0, 1, ..., m_2-1=299$. The shaded area marks time window from which snapshots are used to construct the HODMD model.}
	\label{fig:Amp0p5_md2_step2}
\end{figure}

The extrapolation of $G$ away from the diagonal for $t'=120$ is shown in Figure~\ref{fig:Amp0p5_md2_step2}. Compared to Figure~\ref{fig:Amp0p5_md1_traj2},  the extrapolated $G^{\rm{DMD}}(0; t, 120)$ produced here is much closer to the numerical solution of the KBE than that produced by the extrapolation described in Section~\ref{sec:fixtprime}. 

\begin{figure}[t!]
	\centering
	\includegraphics[width=0.9\textwidth]{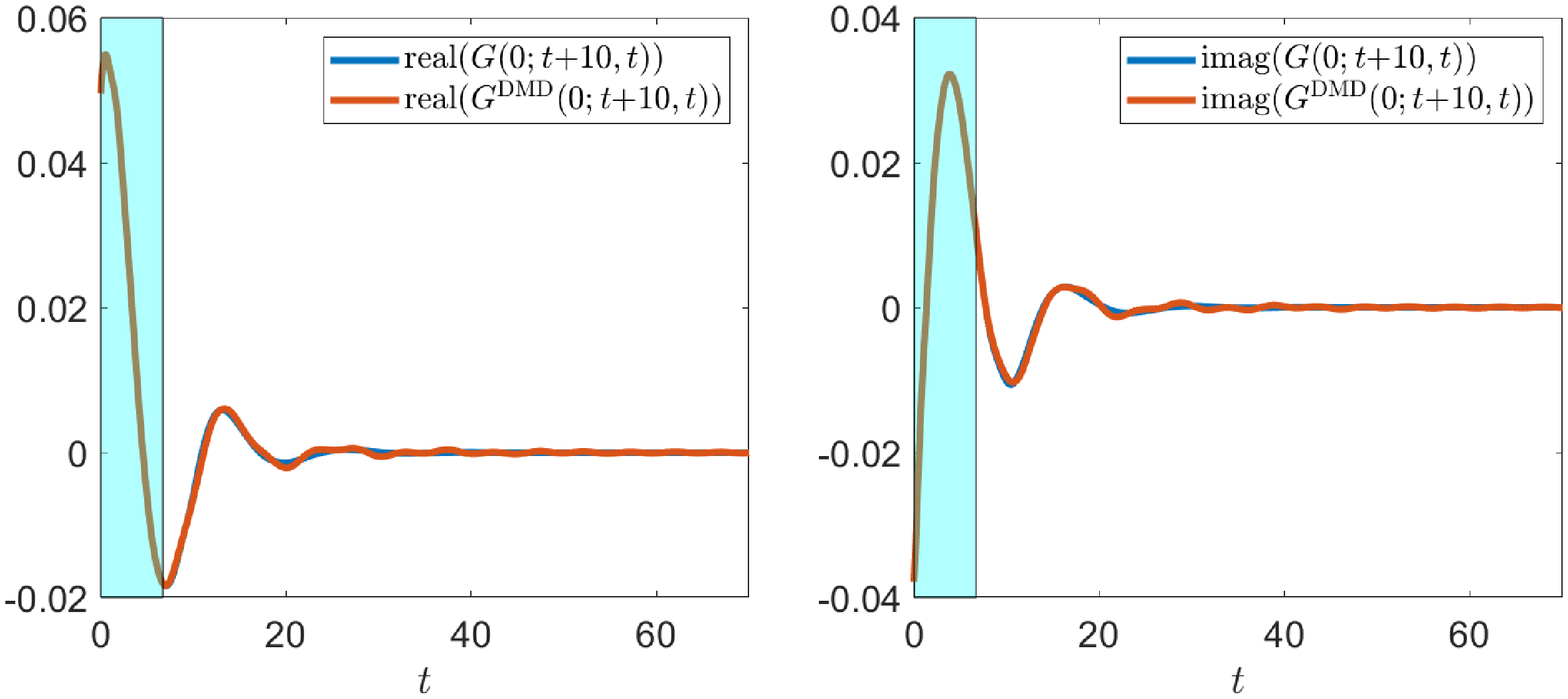}
	\caption{$I=1.5$. A comparison of $G^{\rm{DMD}}(0;t+10,t)$ with $G(0;t+10,t)$. The extrapolation is computed by HODMD(6) with the snapshot matrix $\widetilde{\mathbf{X}}_1$ constructed from $G(0;(p+q)\Delta t, p\Delta t)$ with $p = 0, 1, ..., m_2-1 = 119$ and $q = 0, 1, ..., m_1-1=79$. The shaded area marks time window from which snapshots are used to construct the HODMD model.}
	\label{fig:Amp1p5_md2_step1}
\end{figure}

\begin{figure}[t!]
	\centering
	\includegraphics[width=0.9\textwidth]{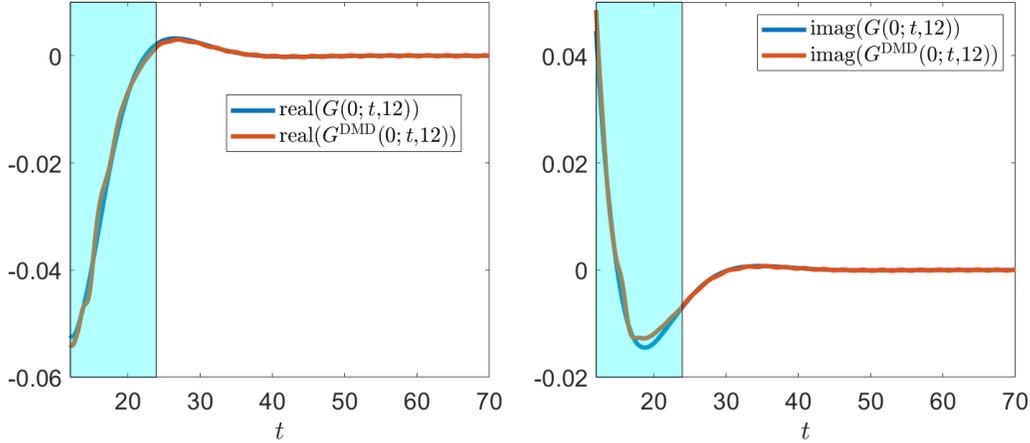}
	\caption{$I=1.5$. A comparison of $G^{\rm{DMD}}(0;t,12)$ with $G(0;t,12)$. The extrapolation is computed by HODMD(5) with the snapshot matrix $\widetilde{\mathbf{X}}_1$ constructed from $G(0;12+(p+q)\Delta t, 12+p\Delta t)$ with $p = 0, 1, ..., n-1 = 29$ and $q = 0, 1, ..., m_2-1=119$. The shaded area marks time window from which snapshots are used to construct the HODMD model.}
	\label{fig:Amp1p5_md2_step2}
\end{figure}

When $I$ is further increased to $I=1.5$, we use HODMD($6$) and HODMD($5$) to extrapolate along the subdiagonal bands and away from the diagonal. In the first step, we take $m_1=80$ snapshots, while in the second step, we take $m_2=120$ snapshots. Extrapolation results at $k=0$ for $G(0; t+10, t)$ from the first step and $G(0; t, 12)$ from the second step are given in Figures~\ref{fig:Amp1p5_md2_step1} and \ref{fig:Amp1p5_md2_step2}, respectively. From Figure~\ref{fig:Amp1p5_md2_step1}, we notice that, although there is not a complete oscillation period in the sampled window, HODMD gives a good approximation to the magnitude and frequency of the oscillation for large $t$ and $t'$. Furthermore, for $t'=12$, which falls out of the sampled window in the first step, the extrapolation from the second step matches well with the numerical solution of KBE. This agreement can be observed from Figure~\ref{fig:Amp1p5_md2_step2}.

\subsection{Spectral function and band structure approximation}
Once $G(k;t,t')$ is available, we can use it to evaluate the spectral function.
The spectral function associated with $G = G_{11} + G_{22}$ at a particular $k$-point $k$ can be computed by the formula
\begin{equation}
	A(T, k, \omega) = {\rm{imag}}\left(\int_{t_0}^Tdt_2\int_{t_0}^Tdt_1s(t_2)s(t_1)e^{i\omega(t_1-t_2)}G(k; t_1, t_2)\right),
\label{eq:atomega}
\end{equation}
where
\begin{equation}
	s(t) = e^{-\frac{(t-(T+\Delta t)/2)^2}{2(1000\Delta t)^2}}.
\end{equation}
We can evaluate
\eqref{eq:atomega} numerically by first performing a discrete Fourier transform with respect to $t_1$ and using the trapezoid rule to evaluate the integral with respect to $t_2$. 

\begin{figure}[t!]
    \quad
    \centering
	\includegraphics[width=0.42\textwidth]{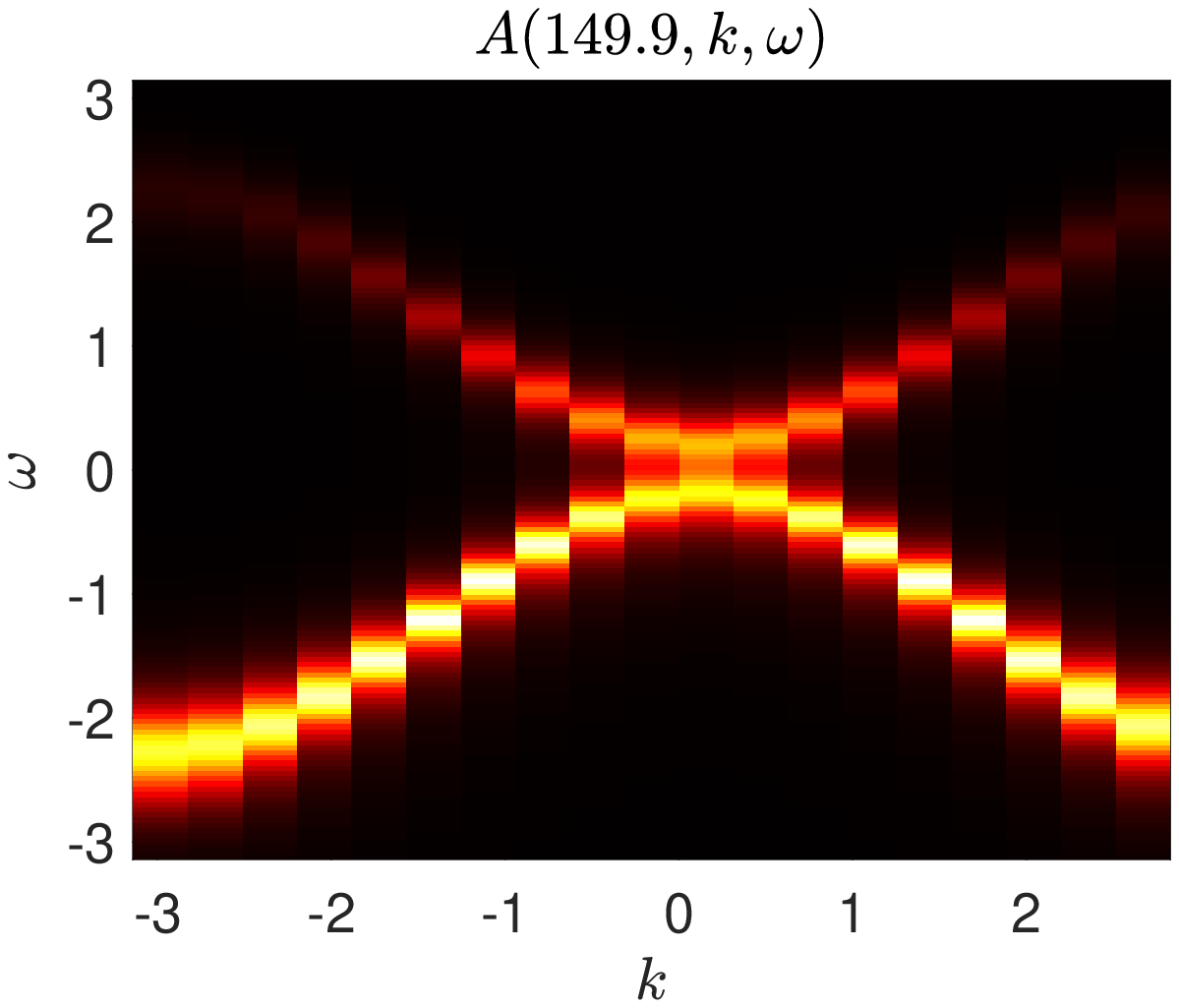}
	\includegraphics[width=0.42\textwidth]{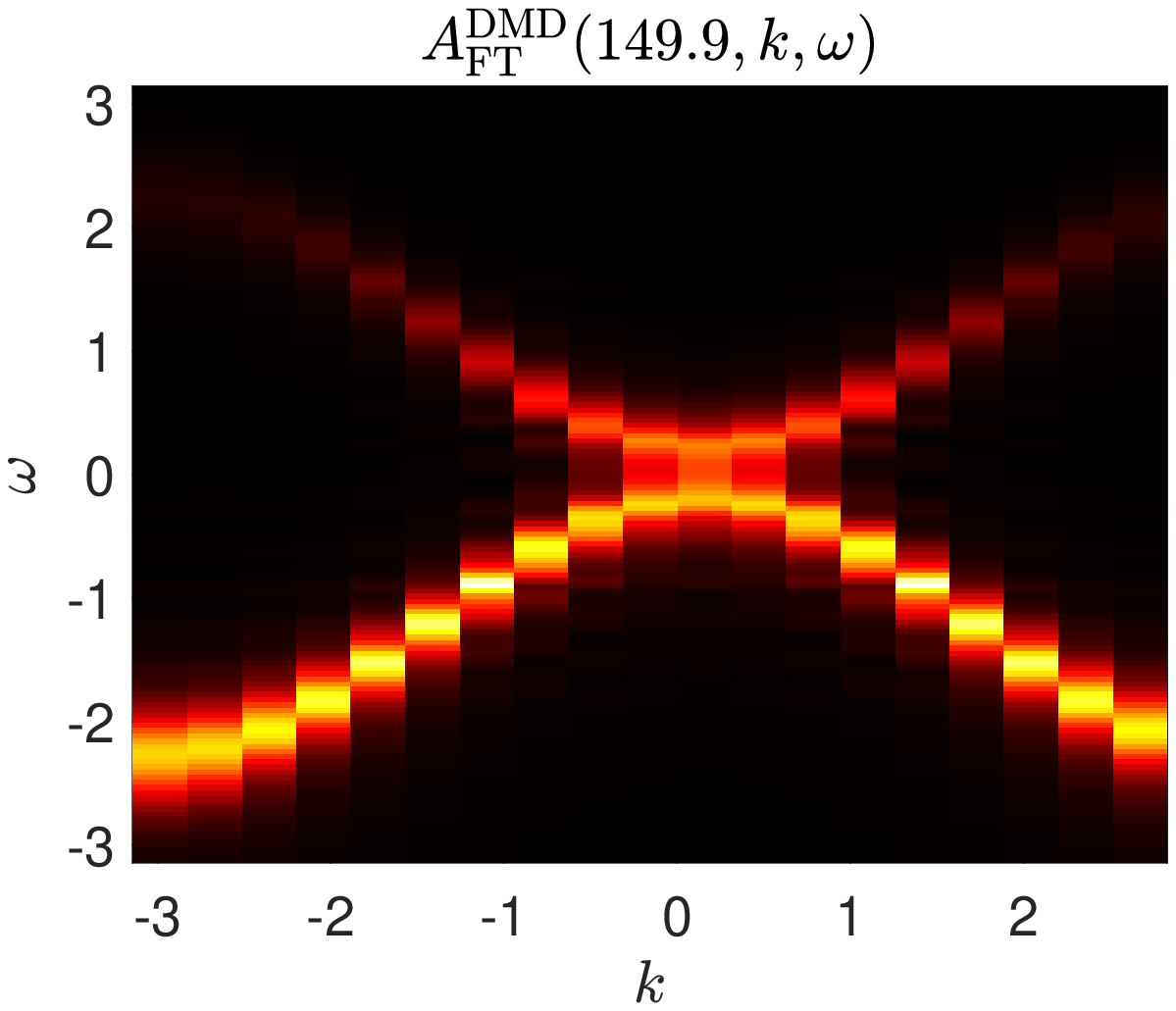}
	\caption{A comparison of band structure constructed from the numerical solution of the KBE for two-band model problem defined by~\eqref{eq:Htotal} and~\eqref{eq:dipole} with $I=1.5$, and a fixed $k$-point DMD extrapolation.}
	\label{fig:Aw_Amp1p5}
\end{figure}

By assembling spectral functions at multiple $k$-points, we can plot the band structure. In Figure~\ref{fig:Aw_Amp1p5}, we compare the band structures $A(149.9,k,\omega)$, plotted as a heatmap, obtained from the numerical solution of the KBE as well as the extrapolated two-time Green's function $G^{\mathrm{DMD}}(t,t')$ for $I = 1.5$. The extrapolated Green's function is obtained by using the fixed timeline (FT) scheme discussed in Section~\ref{sec:fixk}. 
The fixed $k$-point (FK) extrapolation yields a nearly identical result and is not shown here for simplicity. We can clearly see that the band structure constructed from the $G^{\mathrm{DMD}}(t,t')$ is nearly indistinguishable from that constructed from the numerical solution of the KBE. Similar results are observed for band structures constructed from external fields with different intensities, i.e. $I=0.001$ and $I=0.5$.

%

\begin{figure}[t!]
	\centering
	\includegraphics[width=0.42\textwidth]{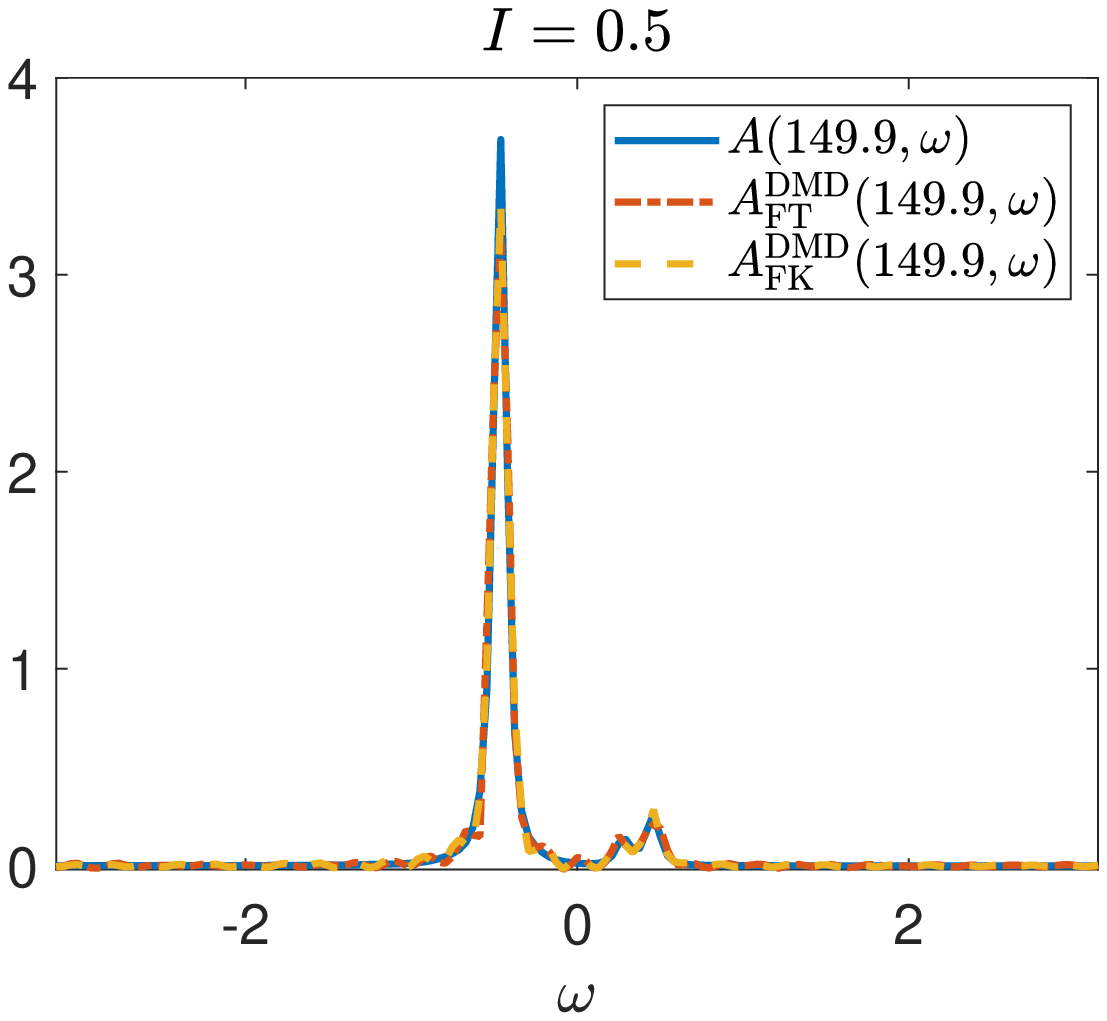}
	\includegraphics[width=0.42\textwidth]{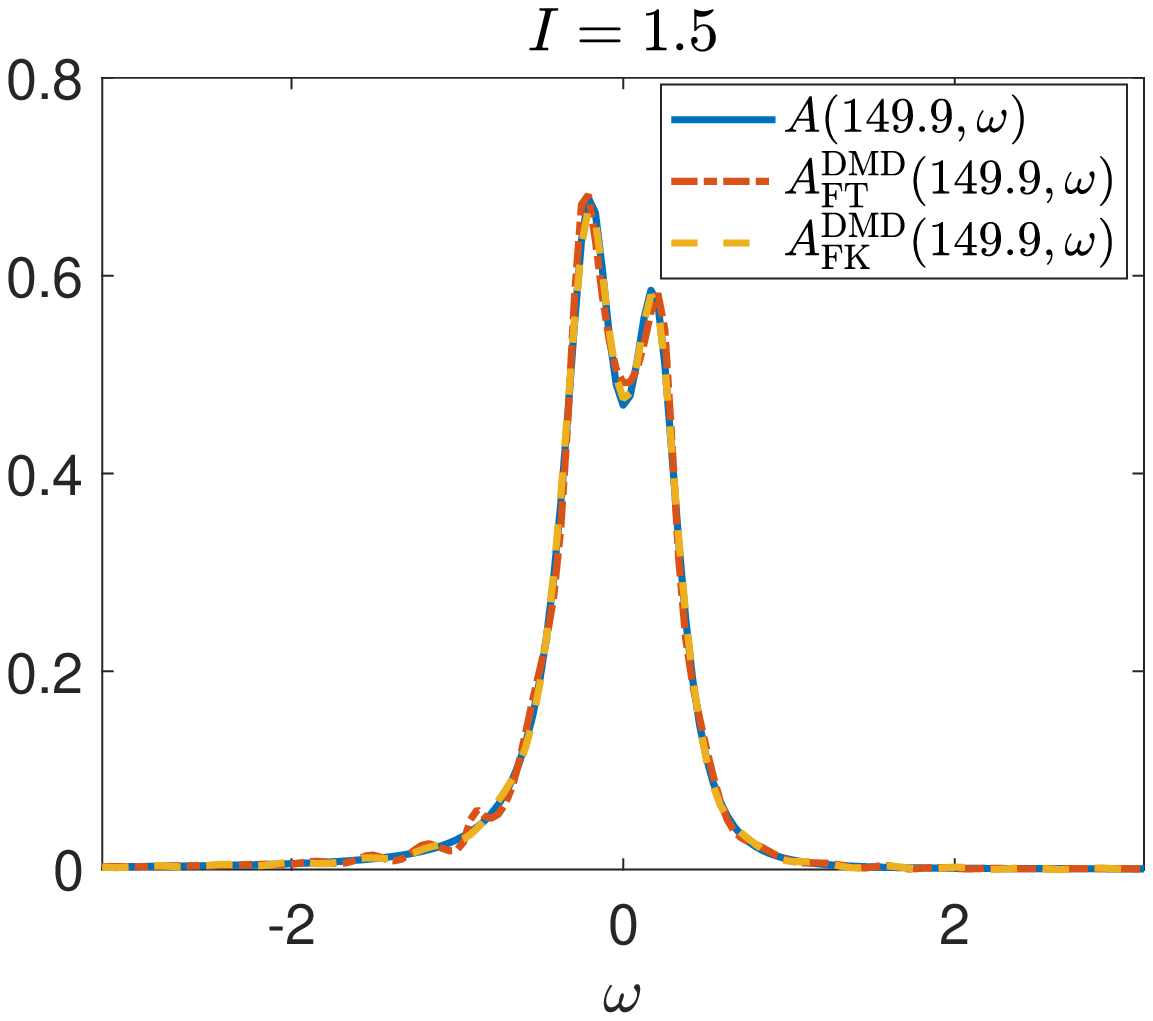}
	\caption{Comparisons of the spectral functions at $k=0$ when the coupling density $I=0.5$ and $I=1.5$.}
	\label{fig:spectral}
\end{figure}

In Figure~\ref{fig:spectral}, we take a closer look at the spectral functions at $k=0$ (the central slice in Figure~\ref{fig:Aw_Amp1p5}. We compare $A_{\rm{FT}}^{\rm{DMD}}$ with $A_{\rm{FK}}^{\rm{DMD}}$, and observe that they both match well with the spectral function obtained from the numerical solution of the KBE for $I=0.5$ and $I=1.5$.  In particular, all major peaks of the spectral function are captured accurately.  Remarkably, the small discrepancy between the extrapolated and the computed Green's functions shown in Figure~\ref{fig:Amp0p5_md1_traj2} has very little effect on the accuracy of the spectral function. 

To determine how the accuracy of HODMD extrapolation is affected by the number of snapshots we need to collect in both FT and FK based HODMD methods, we examine the RMS error of the spectral function for a particular choice of $m_1$ and $m_2$, defined as
\begin{equation}\label{eq:err2}
    {\rm{RMS}}_{\rm{method}}(m_1, m_2) = \frac{2\pi}{T+\Delta t}\sqrt{\sum_{\ell=0}^{T/\Delta t}\left(A(T, \omega_\ell)-A_{\rm{method}}^{\rm{DMD}}(T, \omega_\ell)\right)^2}, \quad j=1, 2,
\end{equation}
where method is either FT or FK, and
\begin{equation}
    \omega_\ell = -\frac{\pi}{\Delta t}+\frac{2\ell\pi}{T+\Delta t}, \quad \ell = 0, ..., \frac{T}{\Delta t}=1499.
\end{equation}

\begin{figure}[t!]
    \qquad
	\centering
	\includegraphics[width=0.9\textwidth]{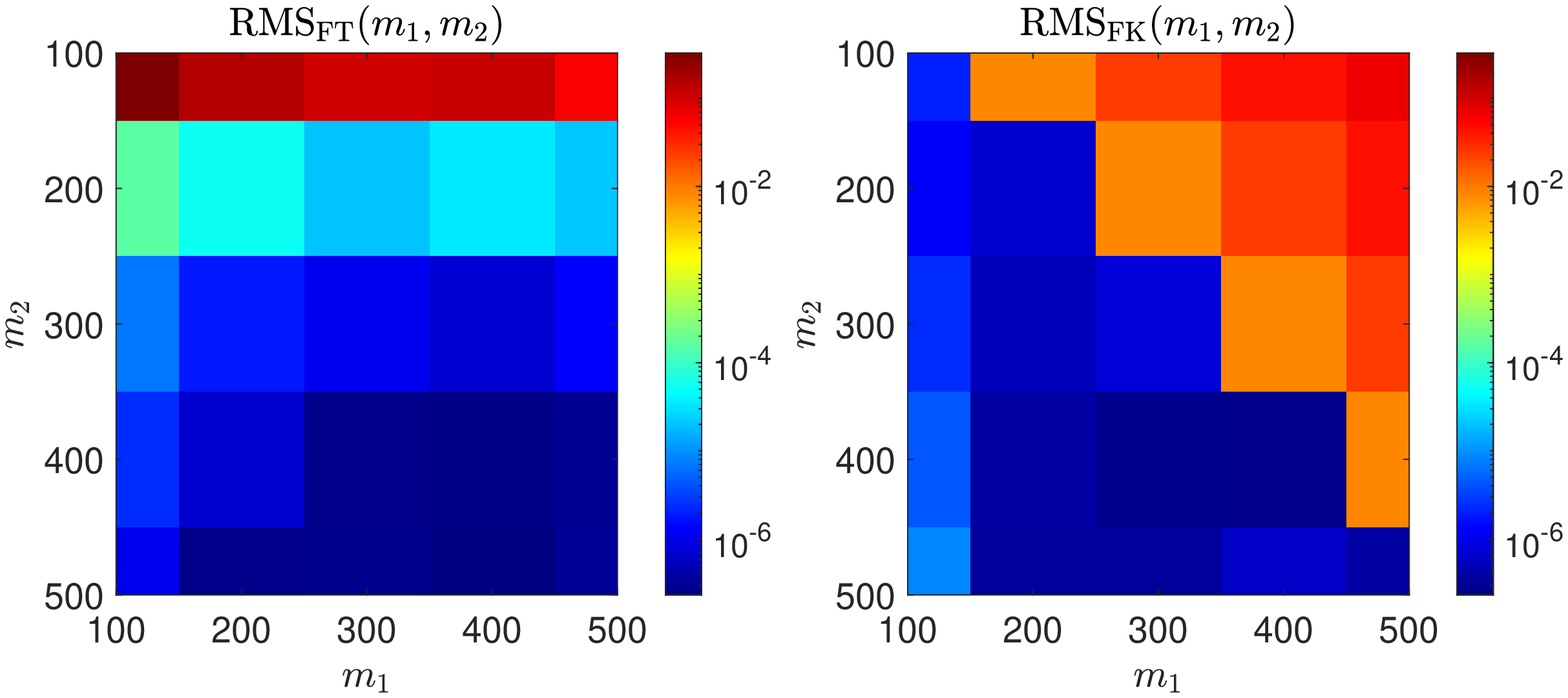}
	\caption{$I=0.001$. The $l_2$-error \eqref{eq:err2} for the spectral function computed from the two HODMD extrapolated Green's function introduced in Sections~\ref{sec:fixt} and~\ref{sec:fixk} respectively.}
	\label{fig:err2_Amp0p001_cmp}
\end{figure}

\begin{figure}[t!]
    \qquad
	\centering
	\includegraphics[width=0.9\textwidth]{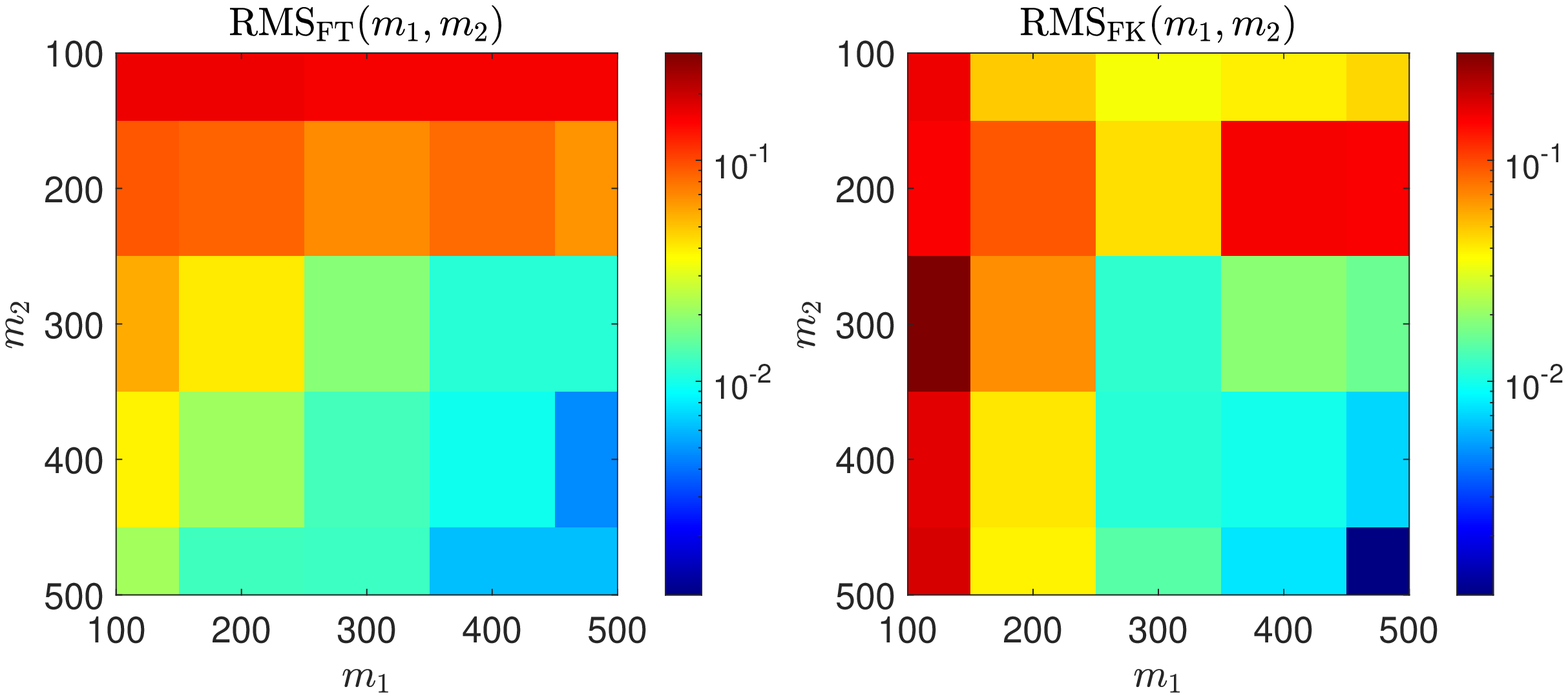}
	\caption{$I=0.5$. The $l_2$-error \eqref{eq:err2} for the spectral function computed from the two HODMD extrapolated Green's function introduced in Sections~\ref{sec:fixt} and~\ref{sec:fixk} respectively}
	\label{fig:err2_Amp0p5_cmp}
\end{figure}

\begin{figure}[t!]
    \qquad
	\centering
    \includegraphics[width=0.9\textwidth]{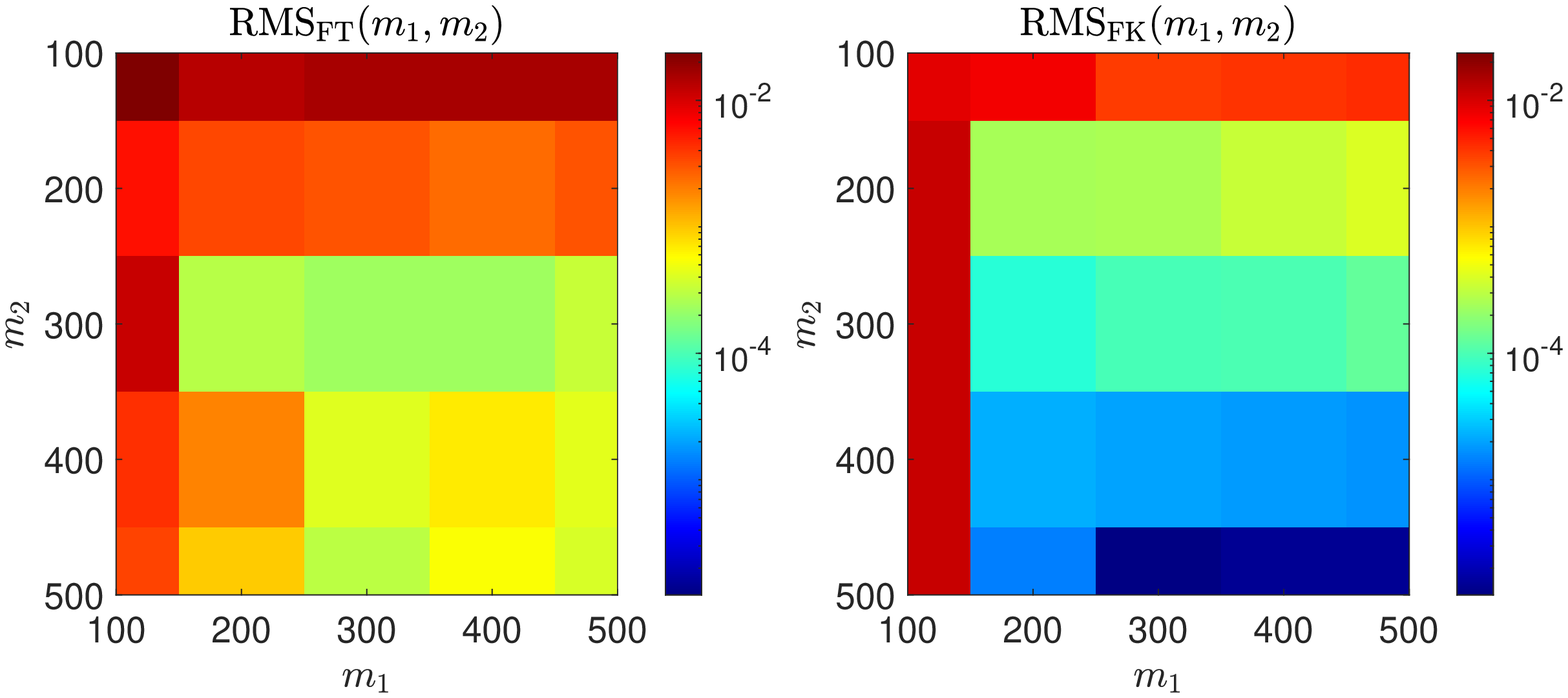}
	\caption{$I=1.5$. The $l_2$-error \eqref{eq:err2} for the spectral function computed from the two HODMD extrapolated Green's function introduced in Sections~\ref{sec:fixt} and~\ref{sec:fixk} respectively}
	\label{fig:err2_Amp1p5_cmp}
\end{figure}

We plot ${\rm{RMS}}_{\rm{FT}}(m_1,m_2)$ and ${\rm{RMS}}_{\rm{FK}}(m_1,m_2)$ for several values of $m_1$ and $m_2$ between 100 and 500 in Figures~\ref{fig:err2_Amp0p001_cmp}-\ref{fig:err2_Amp1p5_cmp} for $I = 0.001$, $I= 0.5$ and $I=1.5$ respectively. Overall, we can observe that the RMS error decreases as $m_1$ and $m_2$ increase.    
When a sufficiently large set of $m_1$ and $m_2$ values are chosen, we observe that 
$A_{\rm{FK}}$ tends to be more accurate than $A_{\rm{FT}}$. 
This observation is also consistent with what we observed in Figures~\ref{fig:Amp0p5_md1_traj2} and \ref{fig:Amp0p5_md2_step2}, where the fixed $k$-point based HODMD extrapolation outperforms the fixed timeline based extrapolation.

We observe in Figure~\ref{fig:err2_Amp0p001_cmp} that ${\rm{RMS}}_{\rm{FK}}(m_1,m_2)$ is 
relatively large when $m_2 < m_1$ regardless how large $m_2$ is. We are not clear at 
the moment why $m_2$ must be larger than $m_1$ in this case.


\subsection{Computational efficiency}

In this section, we report the performance gain achieved by using DMD to extrapolate the non-equilibrium Green's function from the solution of the KBE within a small two-time window. We use the numerical method presented in~\cite{balzer2007nonequilibrium} to solve the KBE associated with two-band model problem with second-Born approximation to the self-energy. The computation is carried out on Cori KNL computer maintained at NERSC. Message Passing Interface (MPI) is used to distribute Green's functions at different $k$-points among different MPI ranks. These Green's functions can be updated in parallel. However, the evaluation of the self-energy and the integral term of the KBE requires global communication. For $20$ $k$-points distributed among $20$ MPI ranks, and wall clock minutes required to perform $500$, $1000$ and $2000$ steps of time evolution respectively are listed in Table \ref{tb:comput}. 
\begin{table}[hb!]
	\centering
	\renewcommand{\arraystretch}{1.2}
	\begin{tabular}{|c|c|c|c|}
		\hline
		Number of time steps & $500$ & $1000$ & $2000$ \\
		\hline
		wall clock time (min) & $13$ & $97$ & $821$ \\
		\hline
	\end{tabular}
    \caption{The wallclock time use to perform numerical time evolution of the KBE on 20 MPI ranks.}
    \label{tb:comput}
\end{table}

From this table, we can clearly see the $O(t^3)$ scaling of the computational cost for solving the KBE. In fact, for this relatively small problem, performing $2000$ steps of time evolution takes nearly 14 wall clock hours on 20 MPI ranks. 

The use of DMD can significantly reduce the computational time and memory cost.  As we indicated earlier, for $I=0.001$ and $I=1.5$, we can use DMD to extrapolate the entire Green's function from the numerical solution of the KBE within a small time window by performing only $500$ steps of time evolution. 

The wall clock time used to perform a fixed timeline (FT) DMD extrapolation on $20$ MPI ranks is less than one minute. On the other hand, a fixed $k$-point DMD extrapolation can be done within $10$ seconds.  These time costs are negligible compared to the time required to solve the KBE. As a result, the use of DMD can speedup the entire computation by a factor of $821/14\approx 59$.

For $I=0.5$, we need to perform $1000$ steps of time evolution before applying DMD extrapolation. The DMD extrapolation can still be completed in less one minute which results in a speedup of $821/98\approx 8$.

Moreover, by reducing the number of time steps used to solve the KBE, we can reduce the memory cost for storing each self-energy function from $5$GB to $320$MB.
\section{Conclusion} \label{sec:conclude}

In this paper, we applied dynamic mode decomposition (DMD), which is a data-driven model order reduction technique, to predict the long-time dynamics of the two-time nonequilibrium Green's function from the numerical solutions of the Kadanoff-Baym equations (KBEs) within a small time window. While the original DMD is applicable to one-time dynamics only, we successfully used it to extrapolate two-time nonlinear dynamics by decomposing the two-time Green's function into several one-time functions in two different time directions. We presented two different time partitioning schemes and compared their effectiveness through numerical examples. We also presented a scheme in which one of the time variable is treated as a spatial variable at a fixed momentum grid point. Our numerical results show that this scheme can sometimes provide a more accurate prediction of the two-time Green's function in a large two-time window.  We have also demonstrated that the DMD extrapolated Green's function can be used to compute interesting physical observables such as the band structure and the spectral functions of the many-body system accurately. By applying DMD/HODMD, we can significantly reduce the computational cost for computing these quantities because we do not need to solve the KBEs within a large two-time window. The cost of performing DMD and HODMD is negligible compared to the cost of solving the KBEs.



\section*{Acknowledgments}
This work is supported by the Center for Computational Study of Excited-State Phenomena in Energy Materials (C2SEPEM) at the Lawrence Berkeley National Laboratory, which is funded by the U.\,S. Department of Energy, Office of Science, Basic Energy Sciences, Materials Sciences and Engineering Division, under Contract No. DE-AC02-05CH11231, as part of the Computational Materials Sciences Program. 
The authors acknowledge the computational resources of the National Energy Research Scientific Computing (NERSC) center.

\bibliographystyle{abbrv}
\bibliography{ref}

\end{document}